\documentclass[aps,twocolumn,superscriptaddress,nofootinbib, longbibliography]{revtex4-1}
\usepackage[latin9]{inputenc}
\usepackage{color}
\usepackage{float}
\usepackage{amsmath}
\usepackage{amssymb}
\usepackage{graphicx}
\usepackage{esint}

\makeatletter

\floatstyle{ruled}
\newfloat{algorithm}{tbp}{loa}
\providecommand{\algorithmname}{Algorithm}
\floatname{algorithm}{\protect\algorithmname}

\makeatother

\begin{document}

\title{xTRAM: Estimating equilibrium expectations from time-correlated simulation data at multiple thermodynamic states}

\author{Antonia S. J. S. Mey}
\email[Electronic Address: ]{antonia.mey@fu-berlin.de}
\affiliation{Freie Universit\"at Berlin, Arnimallee 6, 14195 Berlin}
\author{Hao Wu}
\email[Electronic Address: ]{hwu@zedat.fu-berlin.de}
\affiliation{Freie Universit\"at Berlin, Arnimallee 6, 14195 Berlin}
\author{Frank No\'e}
\email[Electronic Address: ]{frank.noe@fu-berlin.de}
\thanks{To whom correspondence should be addressed}
\affiliation{Freie Universit\"at Berlin, Arnimallee 6, 14195 Berlin}

\date{01.12.2014}

\begin{abstract}
Computing the equilibrium properties of complex systems, such as free energy differences, is often hampered by rare events in the dynamics. Enhanced sampling methods may be used in order to speed up sampling by, for example, using high temperatures, as in parallel tempering, or simulating with a biasing potential such as in the case of umbrella sampling. The equilibrium properties of the thermodynamic state of interest (e.g., lowest temperature or unbiased potential) can be computed using reweighting estimators such as the weighted histogram analysis method or the multistate Bennett acceptance ratio (MBAR). weighted histogram analysis method and MBAR produce unbiased estimates, the simulation samples from the global equilibria at their respective thermodynamic state--a requirement that can be prohibitively expensive for some simulations such as a large parallel tempering ensemble of an explicitly solvated biomolecule. Here, we introduce the transition-based reweighting analysis method (TRAM)--a class of estimators that exploit ideas from Markov modeling and only require the simulation data to be in local equilibrium within subsets of the configuration space. We formulate the expanded TRAM (xTRAM) estimator that is shown to be asymptotically unbiased and a generalization of MBAR. Using four exemplary systems of varying complexity, we demonstrate the improved convergence (ranging from a twofold improvement to several orders of magnitude) of xTRAM in comparison to a direct counting estimator and MBAR, with respect to the invested simulation effort. Lastly, we introduce a random-swapping simulation protocol that can be used with xTRAM, gaining orders-of-magnitude advantages over simulation protocols that require the constraint of sampling from a global equilibrium.

\end{abstract}

\maketitle

\section{Introduction}

The successful prediction of equilibrium
behavior of complex systems is of critical importance in computational
physics. Often, rare events in the system's dynamics make such estimates
through direct simulations impractical. For this reason, the past
20 years have seen vast progress in computational techniques used
for efficient sampling of rare events systems with complex energy
landscapes. These developments were in particular driven by the study
of systems such as spin glasses~\cite{Katzberger2001,Bittner2006},
quantum frustrated spin systems~\cite{Melko2007,Candia2001}, QCD~\cite{Ilgenfritz2002,Burgio2007,Joo1999}
and molecular dynamics (MD) of biomolecules~\cite{Hansmann1997,Sugita1999}.

Commonly used approaches are generalized ensemble
methods such as simulated tempering (ST)~\cite{Marinari1992}, parallel
tempering (PT)~\cite{Hukushima1996,Hansmann1997,Sugita1999} or replica
exchange molecular dynamics (REMD)~\cite{Greyer1991}. Generalized
ensemble methods can greatly improve the convergence over direct simulation
for systems with high energy barriers but relatively few degrees of
freedom~\cite{Sindhikara2010,Park2008,Abraham2008}. The speed of
convergence depends on the overlap of energy distributions between
adjacent temperatures, and thus efforts have been made in choosing optimal temperature distributions~\cite{ZhengLevy_PNAS07_REMDmodel,Rosta2009,Rosta2010,Hansmann1997}.
Unfortunately, the number of replicas needed to fill the relevant
range of temperatures, increases with the number of degrees of freedom
of the system, and produce expensive computational requirements for
many-body systems such as explicitly solvated molecules. 

Once a multi-ensemble simulation was generated, there are different estimator options 
that can be used for the analysis of the simulation data in order to extract information 
such as free energy differences between conformations of interest. The simplest estimator
is to bin the simulation data (in a single order parameter or using clusters of a high-dimensional parameter space) 
at the temperature of interest and count the number of occurrences of each of the discrete states. 
We will refer to this estimation method as the direct counting estimator.
An improvement over direct counting of single temperature histograms, is the weighted histogram analysis method (WHAM)~\cite{FerrenbergSwendsen_PRL89_WHAM,Kumar1992}. WHAM makes use of
data from all simulated temperatures by reweighting them to the target
temperature \emph{via} the Boltzmann
density. The traditional WHAM formulation in multi-temperature ensembles
requires to discretize the system's potential energy in order to formulate
a reweighting factor for each energy bin~\cite{GallicchioLevy_JPCB05_TWHAM}.
A formulation of WHAM that avoids the potential-energy binning is
given in~\cite{SouailleRoux_CPC01_WHAM}. Ref.~\cite{Shirts2008}
generalizes this concept and derives the multistate Bennett acceptance
ratio (MBAR) that produces statistically optimal estimates of equilibrium
properties given a set of equilibrium data.

All of the above estimators assume that the data are
sampled from global equilibrium, i.e. simulations at all temperatures
have entirely decorrelated from their starting configurations. This
often requires discarding large amounts of data for producing unbiased
estimates. 
This can lead to very long simulation times  in
order to obtain an uncorrelated sample.

Here, we combine ideas from reweighting
estimators~\cite{Bennett_JCP76_BAR} and Markov model theory~\cite{Schutte1999,Singhal2005,Chodera2006,NoeHorenkeSchutteSmith_JCP07_Metastability,Chodera2007a,Prinz2011}
in order to derive a transition-based reweighting analysis method
(TRAM) for estimating equilibrium properties from simulation data
at multiple thermodynamic states. TRAM differs from established 
reweighting estimators such as direct counting, WHAM or MBAR, in that it uses
conditional transition probabilities between the discrete states or
bins and therefore does not require the underlying data to be in global equilibrium. 
Thus, TRAM can achieve unbiased equilibrium estimates for data that is not
yet in global equilibrium, allowing accurate estimates to be obtained with sometimes orders of magnitude smaller simulation 
times compared to established estimators. 

Markov models are usually used for predicting long term kinetics from short time simulation data~\cite{Noe2009},
requiring the use of sufficiently long lag times when computing the conditional transition probabilities~\cite{Prinz2011}. 
Therefore, extracting kinetics from generalized ensemble simulations is difficult. If desired, one either has to limit the lag time to the 
short contiguous simulations times \cite{Sriraman2005}, or one has to reweight entire trajectories \cite{Chodera2011,PrinzEtAl_JCP11_Reweighting}.
However, a transition matrix can be used to estimate the equilibrium distribution of a system without requiring log lag times~\cite{Prinz2011}.
At a given temperature $T^{I}$, the corresponding transition matrix $\mathbf{P}^{I}$ provides an estimate of the equilibrium distribution $\boldsymbol{\pi}^{I}$ as its dominant eigenvector. However, in order to exploit the existence of high temperatures in the simulation,
an estimator must be constructed that connects the different temperatures in a rigorous
way. This leads to the proposal of a transition-based reweighting
analysis method (TRAM). 

TRAM can also be employed to get estimates from multiple biased simulations, such as umbrella sampling~\cite{Torrie_JCompPhys23_187} or metadynamics~\cite{LaioParrinello_PNAS99_12562}, although we will here focus on applications using multi-temperature ensembles.
In general, by TRAM we refer to a class of estimators with the following behavior:

\begin{enumerate}
\item Given simulations at different thermodynamic states
$I=1,...,m$ (temperatures, bias potentials, ...), and a configuration-space
discretization into discrete states $i=1,...,n$ (binning along an order
parameter or clustering of a high-dimensional space), 
harvest the following statistics:

\begin{enumerate}
\item At each thermodynamic state $I$, the number of
transitions  $c_{ij}^{I}$ observed between configuration states $i$, $j$
at a time lag $\tau$ (here usually the data storage interval).
\item For each sample $\mathbf{x}$ along the trajectory, with thermodynamic
state $I$ and configuration state $i$ the probability weight $\mu^{J}(\mathbf{x})$ this sample $\mathbf{x}$ would be attributed in all other thermodynamic states
$J=1,...,m$, $\mu^{J}(\mathbf{x})$.
\end{enumerate}
\item Compute an optimal estimate of the equilibrium probability
$\pi_{i}^{I}$ for all configuration states $i$ at all thermodynamic
states $I$.
\end{enumerate}

With the help of the equilibrium probabilities $\pi_{i}^{I}$, other equilibrium
expectations can be computed. Due to property (1a), TRAM is a ``transition-based'' estimator
rather than a histogram method. Due to property (1b), TRAM is also a ``reweighting''
estimator. TRAM estimators do not depend on actual temperature-transitions
in the generalized ensemble. Rather, all configurations visited during
the simulation will be used to estimate transition probabilities between
thermodynamic states. 

Different implementations of TRAM estimators and
formulations of their optimality may be possible, we therefore consider
TRAM to be a class of estimators rather than a unique method. In this
paper we propose a TRAM estimator which formally constructs an expanded
$(mn\times mn)$ transition matrix in the joint space of all $m$
thermodynamic states and $n$ configuration states. Therefore, the
present estimator is called expanded TRAM (xTRAM). The stationary
eigenvector of the xTRAM transition matrix contains the equilibrium
probabilities at all thermodynamic states. 

While simulation protocols such as ST, PT and REMD
require a strong overlap of energy distributions between neighboring
temperatures to be efficient for sampling, this is much less relevant
for the usefulness of TRAM estimators as the reweighting factors are
useful information even when the transition probabilities between
thermodynamic states are small. It is thus tempting to design new
simulation protocols that achieve more efficient sampling by sacrificing
the asymptotic global equilibrium property achieved by ST, PT and
REMD, but can still yield unbiased estimates of equilibrium probabilities
when used in conjunction with TRAM estimators. In this paper we make
a first attempt to this end and propose the random swapping (RS) simulation
method. RS achieves rapid mixing between a set of replicas that would
be too sparsely distributed for ST, PT or REMD because it exchanges
without Metropolis-Hastings acceptance step. The associated violation
of global equilibrium can be approximately recovered by xTRAM because
local equilibrium is guaranteed by adjusting lag times $\tau$ during the estimation accordingly. 

xTRAM is shown to be asymptotically
unbiased. Moreover, we show that xTRAM is a generalization of MBAR
which converge to identical estimators for the free energy differences
between thermodynamic states in the limit of global equilibrium, and
to identical estimators for general equilibrium expectations in the
limit of global equilibrium and large statistics.

Using xTRAM and in particular with the combination
of xTRAM and RS, estimates of equilibrium properties of complex dynamical
systems can be obtained with orders of magnitude fewer simulation
data than that required by conventional estimators.
We illustrate the performance of TRAM, MBAR and direct counting
on two bistable model potentials and two explicitly solvated peptides simulated with Molecular Dynamics simulations.

\section{Theory and Methods}

\subsection{Scope}

A configuration $\mathbf{x}$ is a point in configuration
space $\Omega$ containing all quantities characterizing the instantaneous
state of the system, such as particle positions or spins, system volume
(in constant-pressure ensembles), and particle numbers (in ensembles
of constant chemical potential). 

We consider a set of simulation trajectories, each sampling from $\Omega$, 
at a given \emph{thermodynamic state} $I$. A thermodynamic state, here denoted
as capital superscript letters $I$, $J$, $K$, is characterized
by its thermodynamic variables, such as temperature, pressure or chemical
potential. The dynamics are assumed to fulfill microscopic detailed
balance at their respective thermodynamic states.

We consider $\Omega$ to be partitioned into subsets
$S_{i}$ such that $\Omega=\bigcup_{i=1}^{n}S_{i}$. We will subsequently
refer to subsets $S_{i}$ as \emph{configuration
states} and index them by small subscript letters
$i,j,k$. This discretization serves to distinguish
the states that are relevant to the analyst. As such it may consist
of a fine discretization of an order parameter of interest (e.g. magnetization
in an Ising model), or a Voronoi partition obtained from clustering
molecular dynamics data, as is frequently used for the construction
of Markov models. 

TRAM estimators will use statistics from transitions
among configuration states, but also exploit the fact that the statistical
weight of a configuration $\mathbf{x}$ can be reweighted between thermodynamic
states. Consider the following two examples: 
\begin{enumerate}
\item In PT or REMD simulations, the weighting occurs
through the different temperatures: Given a configuration $\mathbf{x}$
with potential energy $U(\mathbf{x})$, the statistical weight at
temperature $T^{I}$ is proportional to $\mathrm{e}^{-u^{I}(\mathbf{x})}$
using the reduced potential energy
\begin{equation}
u^{I}(\mathbf{x})=\frac{U(\mathbf{x})}{k_{B}T^{I}},
\end{equation}
with Boltzmann constant $k_{B}$. 
\item When the simulation setup contains multiple biased
simulations, such as in umbrella sampling or metadynamics, there is
usually a unique temperature $T$, but different potentials $U^{I}(\mathbf{x})=U(\mathbf{x})+B^{I}(\mathbf{x})$
are employed where $B^{I}(\mathbf{x})$ is the $I$th bias potential.
Then the statistical weights in each of these potentials is given
by $\mathrm{e}^{-u^{I}(\mathbf{x})}$ with:
\begin{equation}
u^{I}(\mathbf{x})=\frac{U(\mathbf{x})+B^{I}(\mathbf{x})}{k_{B}T}.
\end{equation}

\end{enumerate}
We generalize this concept following the example
of~\cite{Shirts2008}. In a thermodynamic state $I$, defined by a
particular combination of the potential energy function $U^{I}$,
pressure $p^{I}$, chemical potentials $\mu_{s}^{I}$ of chemical
species $s$ and temperature $T^{I}$, our system has a \emph{reduced
potential }defined by: 
\begin{equation}
u^{I}(\mathbf{x})=\frac{U^{I}(\mathbf{x})+p^{I}V(\mathbf{x})+\sum_{s}\mu_{s}^{I}N_{s}(\mathbf{x})}{k_{B}T^{I}}.
\end{equation}
Here, $V(\mathbf{x})$ is the volume of the system in configuration
$\mathbf{x}$ and $N_{s}(\mathbf{x})$ counts the particle numbers
of species $s$ at configuration $\mathbf{x}$. The probability density
of configuration $\mathbf{x}$ can, for any arbitrarily chosen thermodynamic
state, be expressed as:
\begin{equation}
\rho^{I}(\mathbf{x})=\frac{1}{Z^{I}}e^{-u^{I}(\mathbf{x})},\label{eq:generalized_statistical_weight}
\end{equation}
where $Z^{I}$ is the partition function of thermodynamic state $I$:
\begin{equation}
Z^{I}=\int_{\Omega}d\mathbf{x}\: e^{-u^{I}(\mathbf{x})}.
\end{equation}
The partition function of configuration state $i$ at thermodynamic
state $I$ is:
\begin{equation}
Z_{i}^{I}=\int_{S_{i}}d\mathbf{x}\: e^{-u^{I}(\mathbf{x})}.
\end{equation}

\subsection{Aims}

Next, we will define the quantities we would like to estimate using TRAM. The \emph{reduced free energy}
of thermodynamic state $I$, $f^{I}$, and the reduced free energy
of configuration state $i$ at thermodynamic state $I$,$f_{i}^{I}$, here termed
the \emph{configuration free energy}
are defined as:
\begin{align}
f^{I} & :=-\ln Z^{I}\label{eq:def_reduced_free_energy}\\
f_{i}^{I} & :=-\ln Z_{i}^{I}.\label{eq:def_configuration_free_energy}
\end{align}
The \emph{equilibrium probability }\textcolor{black}{of
configuration state $i$, given that the system is at thermodynamic
state $I$ is:
\begin{equation}
\pi_{i}^{I}:=\frac{Z_{i}^{I}}{Z^{I}}=\mathrm{e}^{f^{I}-f_{i}^{I}}.\label{eq:def_equilibrium_probability}
\end{equation}
Finally, we are interested in computing }\textcolor{black}{\emph{expectation
values}}\textcolor{black}{{} of arbitrary functions of configuration
state $A(\mathbf{x})$:
\begin{equation}
\langle A\rangle_{I}=\int_{\Omega}d\mathbf{x}\:\rho^{I}(\mathbf{x})\: A(\mathbf{x}).\label{eq:def_expectation_A}
\end{equation}
The multistate Bennett acceptance ratio estimator \cite{Shirts2008}
can provide statistically optimal estimates for quantities (\ref{eq:def_reduced_free_energy})-(\ref{eq:def_expectation_A})
when all samples $\mathbf{x}$ used from the set of simulation data
are independently drawn from the global equilibrium distributions
at the respective thermodynamic states. This requirement can induce
a large statistical inefficiency that we will attempt to avoid by
deriving an estimator that does not rely on global equilibrium.}

\subsection{xTRAM}
The expanded TRAM estimator is based on the idea of
constructing a Markov-model like transition process in an expanded
space whose states are defined by combinations of configuration states
and thermodynamic states. An expanded state is the pairing of
thermodynamic state $I$ and configuration state $i$. We use the
convention of ordering all expanded vectors and matrices first in
blocks of equal thermodynamic state, and within each such block by
configuration state. 

At a given thermodynamic state $I$, the matrix
$\mathbf{C}^{I}=(c_{ij}^{I})$ contains the number of transitions
that have been observed in the data between pairs of configuration
states $i$ and $j$. The diagonal matrix $\mathbf{B}^{IJ}=\mathrm{diag}(b_{i}^{IJ})$
contains transition counts for each configuration state $i$ from
thermodynamic state $I$ to $J$ that have not been observed, but
are constructed so as to obey the correct reweighting between thermodynamic
states. The expanded transition count matrix $\mathbf{\tilde{N}}\in\mathbb{R}^{nm\times nm}$
is given by:
\begin{equation}
\mathbf{\tilde{N}}=\left(\begin{array}{cccc}
\mathbf{C}^{1}+\mathbf{B}^{1,1} & \mathbf{B}^{1,2} & \cdots & \mathbf{B}^{1,m}\\
\mathbf{B}^{2,1} & \mathbf{C}^{2}+\mathbf{B}^{2,2} & \ddots & \vdots\\
\vdots & \ddots & \ddots & \mathbf{B}^{m-1,m}\\
\mathbf{B}^{m,1} & \cdots & \mathbf{B}^{m,m-1} & \mathbf{C}^{m}+\mathbf{B}^{m,m}
\end{array}\right).
\end{equation}

$\mathbf{\tilde{N}}$ has a sparse structure given
by diagonal blocks and off-diagonal bands, as indicated below:
\textcolor{black}{
\begin{equation}
\underset{\mathbf{C}^{1},...,\mathbf{C}^{m}}{\underbrace{\begin{array}{cccc}
\blacksquare\\
 & \ddots\\
 &  & \ddots\\
 &  &  & \blacksquare
\end{array}}}+\underset{\mathbf{B}^{11},...,\mathbf{B}^{mm}}{\underbrace{\begin{array}{cccc}
\diagdown & \diagdown & \cdots & \diagdown\\
\diagdown & \diagdown & \ddots & \vdots\\
\vdots & \ddots & \ddots & \diagdown\\
\diagdown & \cdots & \diagdown & \diagdown
\end{array}}}=\underset{\mathbf{\tilde{N}}}{\underbrace{\begin{array}{cccc}
\blacksquare & \diagdown & \cdots & \diagdown\\
\diagdown & \blacksquare & \ddots & \vdots\\
\vdots & \ddots & \ddots & \diagdown\\
\diagdown & \cdots & \diagdown & \blacksquare
\end{array}}}.\label{eq:N-structure}
\end{equation}
}

\begin{figure}

\begin{centering}
\includegraphics[width=1\columnwidth]{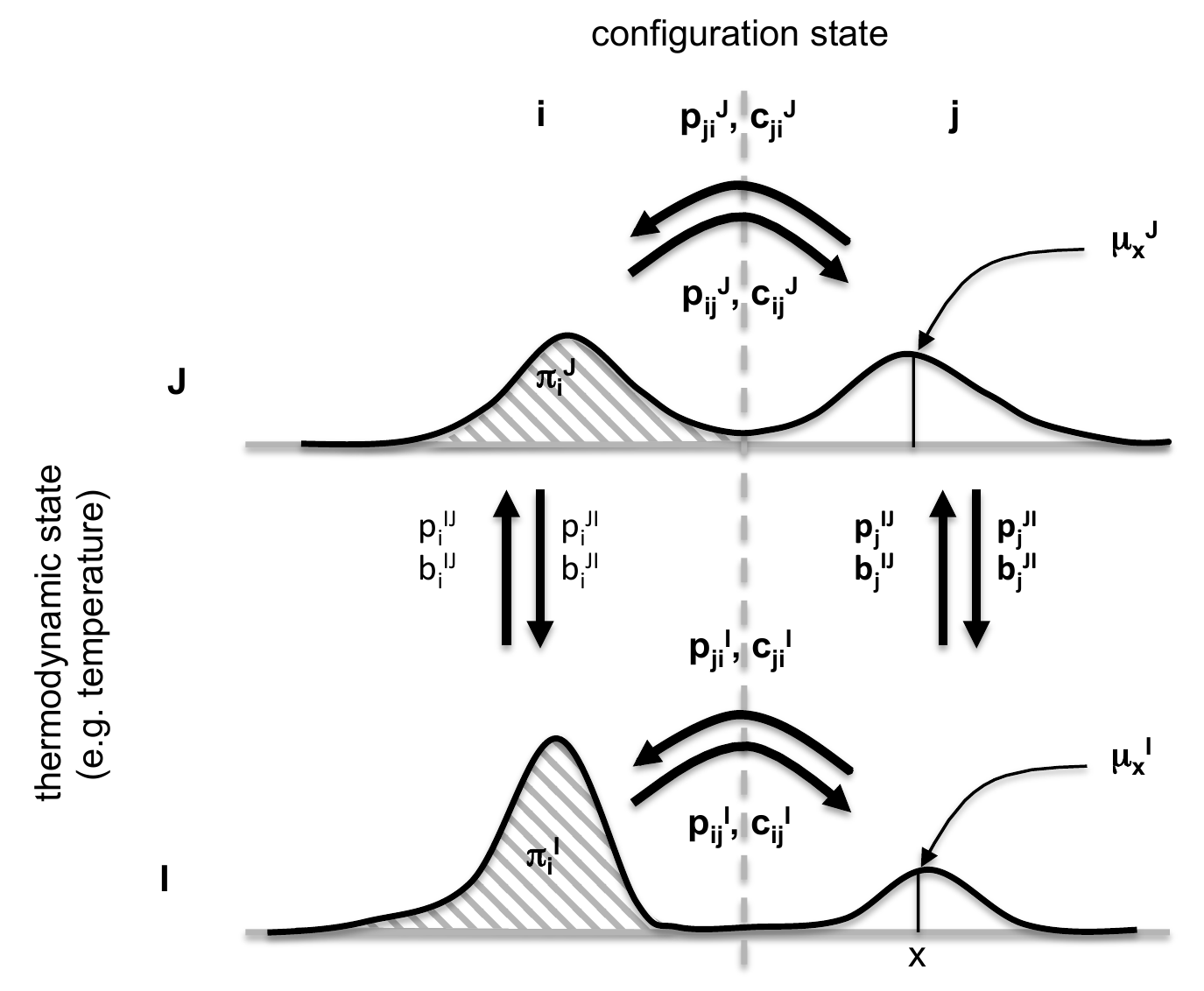}
\par\end{centering}

\caption{Illustration of the expanded transition process in xTRAM and the symbols
used: $c_{ij}^{I}$ and $p_{ij}^{I}$ are transition counts and probabilities
between different configuration states $i$, $j$ at the same thermodynamic
state $I$. $b_{i}^{IJ}$ and $p_{i}^{IJ}$ are transition counts
and probabilities between different thermodynamic states $I$, $J$
at the same configuration state $i$. The latter are constructed such
that reweighting to the equilibrium densities $\mu_{\mathbf{x}}^{I}$,
$\mu_{\mathbf{x}}^{J}$ of configurations $\mathbf{x}$ at different
thermodynamic states $I$, $J$, occurs.}

\end{figure}

The expanded transition matrix $\mathbf{\tilde{P}}$
is defined as the maximum likelihood reversible transition matrix
given $\mathbf{\tilde{N}}$. The key step in xTRAM is to estimate
$\mathbf{\tilde{P}}$ from $\mathbf{\tilde{N}}$ so as to compute
the expanded stationary distribution $\widetilde{\boldsymbol{\pi}}^{\top}=\widetilde{\boldsymbol{\pi}}^{\top}\mathbf{\tilde{P}}$,
which has the structure
\begin{align}
\widetilde{\boldsymbol{\pi}}^{\top} & =(w^{1}\boldsymbol{\pi}^{1},...,w^{m}\boldsymbol{\pi}^{m}),
\end{align}
consisting of subvectors, $\boldsymbol{\pi}^{I}=(\pi_{1}^{I},...,\pi_{n}^{I})$,
each containing the normalized equilibrium probabilities of configuration
states $i$ given that the system is at thermodynamic state $I$.
The weights $w^{I}$, normalized to $\sum_{I}w^{I}=1$, scale all
subvectors such that the expanded equilibrium vector is also normalized,
$\sum_{i}\sum_{I}w^{I}\pi_{i}^{I}=1$.

\paragraph{Data preparation and configuration-state transition
counts}

We process all trajectory data as follows. Each
sample $\mathbf{x}$ occurring in a trajectory at time $t$ is selected
when a successor sample $\mathbf{y}$ at time $t+\tau$ exists
such that the trajectory fragment $\mathbf{x}\rightarrow\mathbf{y}$
was generated using the same dynamics (i.e. without intermediate changes
of the thermodynamic state).

All such samples $\mathbf{x}$ are sorted into sets
$S_{i}^{I}$ according to their configuration state $i$ and thermodynamic
state $I$. The configuration-state transition counts
\[
c_{ij}^{I}=\left|\{(\mathbf{x}\in S_{i}^{I},\mathbf{y}\in S_{j})\}\right|
\]
count the number of transitions from all samples $\mathbf{x}$ in
$S_{i}^{I}$ to target configuration state $j$ ($\mathbf{y}$ itself
can be at any thermodynamic state as long as the dynamics to reach
$\mathbf{y}$ from $\mathbf{x}$ was realized at thermodynamic state
$I$). These counts yield $m$ count matrices $\mathbf{C}^{1},...,\mathbf{C}^{m}$.

We define the total counts $N$, total counts at
thermodynamic state $I$, $N^{I}$, and total counts at thermodynamic
state $I$ and configuration state $i$:

\begin{align}
N_{i}^{I} & :=\sum_{j}c_{ij}^{I}\\
N^{I} & :=\sum_{i}N_{i}^{I}\\
N & :=\sum_{I}N^{I}.
\end{align}
Note that the xTRAM estimations will not depend on the choice of $\tau$
provided that the count matrices $\mathbf{C}^{I}$ are obtained from
simulations that ensure a local equilibrium within each starting state
$S_{i}^{I}$. If this is the case, $\tau$ can be chosen arbitrarily
short, e.g. equal to the interval at which the sampled configuration
are saved. In practice, deviations from local equilibrium can be significant
for certain simulation setups and for poor choices of the discretization,
but can be compensated by using longer lag times (see random swapping
results, below). When local equilibrium is not ensured by the simulation
setup, TRAM estimates should be performed for a series of $\tau$-values,
and then choosing the smallest $\tau$-value for which converged estimates
are obtained.

\paragraph{Thermodynamic-state transition counts}

The elements $I,J$ of the thermodynamic-state count
matrix at configuration state $i$ are constructed by attributing
to each sample $\mathbf{x}$ at thermodynamic state $I$ a single
count that is split to all target thermodynamic states $J$
proportional to the respective probability $p^{IJ}(\mathbf{x})$:
\begin{equation}
b_{i}^{IJ}=\sum_{\mathbf{x}\in S_{i}^{I}}p^{IJ}(\mathbf{x}),\label{eq:N_i^IJ}
\end{equation}
where $p^{IJ}(\mathbf{x})$ is the transition probability to thermodynamic
state $J$ given that the system is at configuration $\mathbf{x}$
and thermodynamic state $I$. In the example of a multi-temperature
simulation, $p^{IJ}(\mathbf{x})$ can be interpreted as the probability
for which a hypothetical simulated tempering trial from temperature $I$ to $J$ 
would be accepted at configuration $\mathbf{x}$. 
In a more general setting, the transition probabilities
between thermodynamic states can be derived from Bennett's acceptance
ratio \cite{Bennett_JCP76_BAR}. From $\sum_{J}p^{IJ}(\mathbf{x})=1$,
it directly follows that $N_{i}^{I}=\sum_{J}b_{i}^{IJ}$. Different
choices for $p^{IJ}(\mathbf{x})$ are possible as long as they respect
detailed balance. 

The statistical weights $w^{I}$ of thermodynamic
states in the expanded ensemble are chosen as the fraction of samples
seen at each thermodynamic state $I$:
\begin{align}
w^{I} & :=\frac{N^{I}}{N}.\label{eq:weights_optimal}
\end{align}
It will be shown later that this choice leads to a statistically optimal
estimator. As an example, consider a replica-exchange simulation,
all replicas $I$ are propagated in parallel and therefore $N^{1} = ... = N^{m}$,
resulting in equal weights $w_{I}=1/m$. When the input data
stems from simulated tempering simulations between different temperatures
$I$, the fraction of time spent at each temperature depends on the
choice of the simulated tempering weights \cite{Marinari1992}, and
could therefore be different. With choices given by Eq. (\ref{eq:weights_optimal}),
the absolute probability of configuration $\mathbf{x}$ in the expanded
ensemble is:
\begin{equation}
\mu^{I}(\mathbf{x})=w^{I}\rho^{I}(\mathbf{x})=\frac{N^{I}}{N}e^{f^{I}-u^{I}(\mathbf{x})}.\label{eq:density_x}
\end{equation}
In order for the thermodynamic-state transition process to sample
from the correct statistical weights, it must fulfill the detailed
balance equations:
\begin{equation}
\mu^{I}(\mathbf{x})\, p^{IJ}(\mathbf{x})=\mu^{J}(\mathbf{x})\, p^{JI}(\mathbf{x}).\label{eq:DB_thermodynamic}
\end{equation}
Various choices for $p^{IJ}(\mathbf{x})$ can be made that meet these
constraints. It will turn out that the statistically optimal choice
is to a thermodynamic state $J$ according to its equilibrium probability
of that state in the expanded ensemble, i.e.:
\begin{equation}
p^{IJ}(\mathbf{x})=\frac{\mu^{J}(\mathbf{x})}{\sum_{K}\mu^{K}(\mathbf{x})}=\frac{N^{J}e^{f^{J}-u^{J}(\mathbf{x})}}{\sum_{K}N^{K}e^{f^{K}-u^{K}(\mathbf{x})}}.\label{eq:p_x^IJ_direct}
\end{equation}
The choices (\ref{eq:density_x}) and (\ref{eq:p_x^IJ_direct}) obviously
fulfill the detailed balance equations (\ref{eq:DB_thermodynamic}).
Using Eq.~(\ref{eq:p_x^IJ_direct}) in an implementation requires
shifting the absolute energy value in order to avoid numerical overflows
when evaluating the exponential (see Appendix of~\cite{Shirts2008}
for a discussion).

An alternative choice for the thermodynamic-state
transition process is the Metropolis rule, which is easier to implement
and produced indistinguishable results compared to choice (\ref{eq:p_x^IJ_direct})
in our applications:
\begin{align}
p^{IJ}(\mathbf{x}) & =\frac{N^I}{N}\min\left\{ 1,\frac{N^{J}\mathrm{e}^{f^{J}-u^{J}(\mathbf{x})}}{N^{I}\mathrm{e}^{f^{I}-u^{I}(\mathbf{x})}}\right\} \:\:\:\:\:\: I\neq J\label{eq:p_x^IJ_Metropolis}\\
p^{II}(\mathbf{x}) & =1-\sum_{J\neq I}p^{IJ}(\mathbf{x})\nonumber .
\end{align}

\paragraph{Free energies:}

In order to compute $p^{IJ}(\mathbf{x})$ in Eq.~(\ref{eq:N_i^IJ}) using Eq. (\ref{eq:p_x^IJ_direct}) or (\ref{eq:p_x^IJ_Metropolis}),
an estimate of the free energies $f^{I}$ is needed. At initialization,
the $f^{I}$'s are estimated using Bennett's acceptance ratio~\cite{Bennett_JCP76_BAR}.
To this end, the thermodynamic states simulated at, are sorted in a
sequence $(1,\:...,I,\: I+1,\:...,\: m)$, e.g. ascending temperatures.
The free energies are then initially set to:
\begin{align}
f^{1} & :=0\\
f^{I+1} & :=f^{I}-\ln\frac{\frac{1}{N^{I}}\sum_{\mathbf{x}\in S^I}\min\{1,\mathrm{e}^{u^{I}(\mathbf{x})-u^{I+1}(\mathbf{x})}\}}{\frac{1}{N^{I+1}}\sum_{\mathbf{x}\in S^{I+1}}\min\{1,\mathrm{e}^{u^{I+1}(\mathbf{x})-u^{I}(\mathbf{x})}\}},
\end{align}
where the sample averages are taken over all samples in a given thermodynamic
state, as denoted by $\mathbf{x}\in S^I$. In subsequent
iterations, we can update the free energies using:
\begin{equation}
f^{I,\,\mathrm{new}}:=f^{I}-\ln\frac{N}{N^{I}}\sum_{i}\tilde{\pi}_{i}^{I}.\label{eq:iteration_f^I}
\end{equation}
The iteration is converged when in the expanded equilibrium distribution
$\boldsymbol{\tilde{\pi}}$ has weights equal according to the target
values $w^{I}$: $\sum_{i}\tilde{\pi}_{i}^{I}=\frac{N^{I}}{N}$ for
all $I$. Eq. (\ref{eq:iteration_f^I}) will adapt $f^{I}$ until
this equilibrium is achieved (see supplementary information for details).

\paragraph{Estimation of the equilibrium distribution}

In every iteration, we obtain a transition
count matrix possessing the sparsity structures sketched in (\ref{eq:N-structure}).
Because the theory is based on a transition matrix fulfilling detailed
balance, we can estimate $\mathbf{\tilde{P}}$ using the reversible
transition matrix estimator described in~\cite{Prinz2011}
which also provides the expanded equilibrium distribution $\boldsymbol{\tilde{\pi}}$
as a by-product.

However, we can derive a simple direct estimator
for $\boldsymbol{\tilde{\pi}}$ without going through $\mathbf{\tilde{P}}$
(see supplementary information I.D). Let $x_{i}^{I}$ be variables
that are iterated to approximate $\pi_{i}^{I}$. Then
iterating the update:
\begin{align}
x_{i}^{I,\:\mathrm{new}} & =\frac{1}{2}\sum_{j=1}^{n}\frac{c_{ij}^{I}+c_{ji}^{I}}{\frac{N_{i}^{I}}{w^{I}\pi_{i}^{I}}+\frac{N_{j}^{I}}{w^{I}\pi_{j}^{I}}}+\frac{1}{2}\sum_{J=1}^{m}\frac{b_{i}^{IJ}+b_{i}^{JI}}{\frac{N_{i}^{I}}{w^{I}\pi_{i}^{I}}+\frac{N_{i}^{J}}{w^{J}\pi_{i}^{J}}}\label{eq:pi_tram_iteration_1}\\
\pi_{i}^{I,\:\mathrm{new}} & =\frac{x_{i}^{I}}{\sum_{j=1}^{n}x_{j}^{I}}.\label{eq:pi_tram_iteration_2}
\end{align}
converges towards the maximum likelihood estimate of $\pi_{i}^{I}$.
The xTRAM estimator is summarized in algorithm~\ref{algo}.

\begin{algorithm}[H]
\begin{enumerate}
\item Compute the largest connected set from the projection
of the multi-temperature trajectory ensembles onto states. All vectors
and matrices are defined on that connected set. For all other states,
$\pi_{i}^{I}$ is set to 0.
\item Initial guess of free energies: set $f_{1}:=0$
and for $I=1,...,m-1$ set:
\[
f^{I+1}:=f^{I}-\ln\frac{\frac{1}{N^{I}}\sum_{\mathbf{x}\in(*,I)}\min\{1,\mathrm{e}^{u^{I}(\mathbf{x})-u^{I+1}(\mathbf{x})}\}}{\frac{1}{N^{I+1}}\sum_{\mathbf{x}\in(*,I+1)}\min\{1,\mathrm{e}^{u^{I+1}(\mathbf{x})-u^{I}(\mathbf{x})}\}}.
\]

\item Compute configuration-state counts $\mathbf{C}^{I}=(c_{ij}^{I})$. $c_{ij}^{I}$ is the number of times a trajectory simulated at
thermodynamic state $I$ was found to be at configuration state $i$ at time $t$, and at state $j$ at time $t+\tau$.
\\
Define $N_{i}^{I}:=\sum_{j}c_{ij}^{I}$, $N^{I}:=\sum_{i}N_{i}^{I}$, $N:=\sum_{I}N^{I}$.

\item Initial guess of equilibrium probabilities:
\[
\pi_{i}^{I}:=\frac{N_{i}^{I}}{N^{I}}.
\]

\item Iterate to convergence of $f^{I}$:

\begin{enumerate}
\item Compute thermodynamic-state counts by
\[
b_{i}^{IJ}:=\sum_{x\in S_{i}^{I}}p^{IJ}(\mathbf{x}),
\]
with $p^{IJ}(\mathbf{x})$ from Eq. (\ref{eq:p_x^IJ_direct}) or (\ref{eq:p_x^IJ_Metropolis}).
\item Iterate to convergence of $\pi_{i}^{I}$ using $w^{I}:=N^{I}/N$:
\begin{align*}
x_{i}^{I,\,\mathrm{new}} & :=\sum_{j}\frac{c_{ij}^{I}+c_{ji}^{I}}{\frac{N_{i}^{I}}{w^{I}\pi_{i}^{I}}+\frac{N_{j}^{I}}{w^{I}\pi_{j}^{I}}}+\sum_{J}\frac{b_{i}^{IJ}+b_{i}^{JI}}{\frac{N_{i}^{I}}{w^{I}\pi_{i}^{I}}+\frac{N_{i}^{J}}{w^{J}\pi_{i}^{J}}}\\
\pi_{i}^{I\,\mathrm{new}} & :=\frac{x_{i}^{I}}{\sum_{I}\sum_{j}x_{j}^{I}}.
\end{align*}

\item \textcolor{black}{Update free energies:
\[
f^{I}:=f^{I}-\ln\frac{N}{N^{I}}\sum_{i}\pi_{i}^{I}.
\]
}
\end{enumerate}
\end{enumerate}
\caption{\label{algo}xTRAM Algorithm for estimating the free energies $f^{I}$
and equilibrium probabilities $\pi_{i}^{I}$.}

\end{algorithm}

\paragraph{Estimation of arbitrary expectation functions}

Now we can derive an efficient estimator of the equilibrium expectation values $\langle A\rangle$
of an arbitrary function $A(\mathbf{x})$, as defined by Eq. (\ref{eq:def_expectation_A}), at an arbitrary thermodynamic
state (possibly not simulated at).
For this we employ Eqs. (14-15) in \cite{Shirts2008}, treating every configuration state at every thermodynamic state 
as a separate MBAR thermodynamic state. We define the weights:
\begin{equation}
g(\mathbf{x})=\frac{\mathrm{e}^{-u(\mathbf{x})}}{\sum_{K}\sum_{i}N_{i}^{K}\mathrm{e}^{f_{i}^{K}-u^{K}(\mathbf{x})}},
\end{equation}
where the configuration free energies can be computed as $f_{i}^{I}=f^{I}-\ln\pi_{i}^{I}$.
As shown in the supplementary information, the expectation values
of an arbitrary function of configuration, $\langle A\rangle$ can thus be estimated as
\begin{equation}
\langle A\rangle=\frac{\sum_{\mathbf{x}}g(\mathbf{x})\, A(\mathbf{x})}{\sum_{\mathbf{x}}g(\mathbf{x})},
\end{equation}
where $\sum_{\mathbf{x}}$ runs over all samples in the data.

\subsection{Asymptotic correctness of the xTRAM estimator}

The exact transition probability between sets $S_{i}$
and $S_{j}$ at thermodynamic state $I$ is given by:
\begin{align}
p_{ij}^{I} & =\frac{1}{\pi_{i}^{I}}\int_{S_{i}}d\mathbf{x}\:\mu^{I}(\mathbf{x})\int_{S_{j}}d\mathbf{y}\: p^{I}(\mathbf{x},\mathbf{y};\,\tau)\label{eq:def_p_ij^I},
\end{align}
where $p^{I}(\mathbf{x},\mathbf{y};\,\tau)$ is the probability density
of the system to be in configuration $\mathbf{y}$ at time $t+\tau$
given that it is in configuration $\mathbf{x}$ at thermodynamic state
$I$ at time $t$. By definition, microscopic detailed balance holds
for the dynamics at thermodynamic state $I$: $\mu^{I}(\mathbf{x})\, p^{I}(\mathbf{x},\mathbf{y};\,\tau)=\mu^{I}(\mathbf{y})\, p^{I}(\mathbf{y},\mathbf{x};\,\tau)$.
Using detailed balance in (\ref{eq:def_p_ij^I}) directly leads to:
\begin{equation}
\pi_{i}^{I}p_{ij}^{I}=\pi_{j}^{I}p_{ji}^{I}.\label{eq:DB_discrete_conf}
\end{equation}
The exact thermodynamic state transition probability from thermodynamic
state $I$ to $J$ at configuration state $i$ is given by:
\begin{equation}
p_{i}^{IJ}=\frac{1}{w^{I}\pi_{i}^{I}}\int_{S_{i}}d\mathbf{x}\:\mu_{\mathbf{x}}^{I}p_{\mathbf{x}}^{IJ}.
\end{equation}
Together with (\ref{eq:DB_thermodynamic}), we have detailed balance
in discrete states:
\begin{equation}
w^{I}\pi_{i}^{I}p_{i}^{IJ}=w^{J}\pi_{i}^{J}p_{i}^{JI}.\label{eq:DB_discrete_therm}
\end{equation}
In the statistical limit $N\rightarrow\infty$, which can be either
realized by trajectories of great length, or by a large number of
short trajectories, our expected transition counts converge to the
following limits:
\begin{align}
\hat{c}_{ij}^{I} & =\lim_{N\rightarrow\infty}c_{ij}^{I}=N_{i}^{I}p_{ij}^{I}\\
\hat{b}_{ij}^{I} & =\lim_{N\rightarrow\infty}b_{ij}^{I}=N_{i}^{I}p_{i}^{IJ}.
\end{align}
Plugging these counts and the reversibility conditions (\ref{eq:DB_discrete_conf},\ref{eq:DB_discrete_therm})
into the estimator of equilibrium probabilities (\ref{eq:pi_tram_iteration_1}),
we obtain the accurate result:
\begin{equation}
x_{i}^{I}=w^{I}\pi_{i}^{I}\label{eq:stat_limit_pi_i^I}.
\end{equation}
Furthermore, in the statistical limit, the Bennett acceptance ratio
initialization (Algorithm \ref{algo}, step 2.) is exact. With result
(\ref{eq:stat_limit_pi_i^I}), this estimate is not changed in Algorithm
\ref{algo}, step 5c.
Thus, the xTRAM estimator converges to unbiased
estimates of all equilibrium properties (\ref{eq:def_reduced_free_energy}-\ref{eq:def_expectation_A})
in the statistical limit.

\subsection{Special cases}

\paragraph{With one thermodynamic state, xTRAM is a Markov
model}

Consider the situation that simulations were conducted at a single thermodynamic
state, such as unbiased molecular dynamics simulations of a macromolecule 
at a fixed temperature $I$. 
The xTRAM count matrix is now an $n\times n$ configuration
state count matrix $\mathbf{C}=(c_{ij})$. 

We only have one free energy $f^{1}=0$. 
Using Eq. (\ref{eq:def_equilibrium_probability}), the configuration free energies 
are given by $f_{i}^{I} = -\ln \pi_{i}$, where $\pi_{i}$ are the estimated
equilibrium probabilities of discrete configuration states $i$. 
These equilibrium probabilities can be obtained by the 
special case of Eq. (\ref{eq:pi_tram_iteration_1},\ref{eq:pi_tram_iteration_2}):
\begin{align}
x_{i}^{\mathrm{new}} & =\sum_{j=1}^{n}\frac{c_{ij}+c_{ji}}{\frac{N_{i}}{\pi_{i}}+\frac{N_{j}}{\pi_{j}}} \label{x_i_MSM} \\
\pi_{i}^{\mathrm{new}} & =\frac{x_{i}}{\sum_{j=1}^{n}x_{j}} \label{pi_i_MSM}.
\end{align}

Eqs. (\ref{x_i_MSM}-\ref{pi_i_MSM}) is the equilibrium probability of the maximum likelihood $n \times n$ 
reversible transition matrix given count matrix $\mathbf{C}$. Therefore, in the single-thermodynamic state
case our estimates are identical to those of a reversible Markov model. 

Standard methods can be used to compute the maximum likelihood reversible transition matrix $\mathbf{P}$
\cite{Prinz2011,Bowman_JCP09_Villin}. However, if we wish to use $\mathbf{P}$ to extract not only stationary but kinetic information,
the lag time $\tau$ used to obtain the count matrix $\mathbf{C}$ must be chosen sufficiently large in order
to obtain an accurate estimate \cite{Prinz2011}.

\paragraph{When all thermodynamic states are in global equilibrium,
xTRAM is identical to MBAR in the estimation of $f^{I}$}

In order to show the relationship between TRAM and
MBAR, we use the TRAM equations (\ref{eq:iteration_f^I},\ref{eq:pi_tram_iteration_1}-\ref{eq:pi_tram_iteration_2}), 
and specialize them using the MBAR assumption that each thermodynamic state is sampled from global equilibrium.
This assumption can be modeled by merging all configuration states to one state. When converged, the
TRAM quantities then fulfill the equations:
\begin{align}
\pi^{I}p^{IJ} & =\frac{b^{IJ}+b^{JI}}{\frac{N^{I}}{\pi^{I}}+\frac{N^{J}}{\pi^{J}}}\\
0 & =-\ln\frac{N}{N^{I}}\pi^{I}.
\end{align}
By combining these equations with (\ref{eq:N_i^IJ}) and (\ref{eq:p_x^IJ_direct})
(see Supplementary Information for details), we obtain:
\begin{equation}
f^{I}=-\ln\sum_{\mathrm{all}\:\mathbf{x}}\frac{e^{-u^{I}(\mathbf{x})}}{\sum_{K}N^{K}e^{f^{K}-u^{K}(\mathbf{x})}},\label{eq:mbar_free_energy}
\end{equation}
which is identical to the MBAR estimator for the reduced free energy
of thermodynamic state $I$ (See Eq. (11) in~\cite{Shirts2008}.

\paragraph{The MBAR and xTRAM estimators of $\pi_{i}^{I}$
are consistent}

Using again the condition that all simulations are
in their respective global equilibria, and tending to the statistical
limit $N\rightarrow\infty$ (see Supplementary Information for details),
we can show that the xTRAM estimate for the equilibrium probabilities
$\pi_{i}^{I}$ can be written as: 
\begin{equation}
\pi_{i}^{I}=\frac{\sum_{\mathbf{x}\in S_{i}}\frac{e^{-u^{I}(\mathbf{x})}}{\sum_{K}N^{K}e^{f^{K}-u^{K}(\mathbf{x})}}}{\sum_{\mathrm{all}\:\mathbf{x}}\frac{e^{-u^{I}(\mathbf{x})}}{\sum_{K}N^{K}e^{f^{K}-u^{K}(\mathbf{x})}}},\label{eq:mbar_equilibrium_distribution}
\end{equation}
which is identical to the MBAR expectation value for $\pi_{i}^{I}$
(to obtain this result, use Eqs. (14-15) in~\cite{Shirts2008}
with the indicator function on set $S_{i}$).

\subsection{Random swapping (RS) simulations}

PT, ST and REMD simulation protocols are constructed
such that they sample from global equilibrium at all temperatures
after a sufficiently long burn-in phase. Global equilibrium is ensured
by constructing appropriate Metropolis acceptance criteria for temperature
swaps. The disadvantage is that to ensure a good mixing rate between
replicas, dense replica spacing is required - a problem that becomes
increasingly difficult for systems with a large number of degrees
of freedom, as is the case of biomolecular simulations of $10^{5}$
or more atoms.

However, due to the use of transition matrices,
xTRAM only requires local equilibrium within the discrete configurational
states rather than global equilibrium - a much weaker requirement.
It is thus tempting to consider using a simulation protocol that is
much more efficient than PT, ST and REMD while sacrificing the property
that it samples from global equilibrium at all temperatures. Such
a protocol would be useful if it is still possible to recover the
correct stationary probabilities using xTRAM. One can consider the
simple random swapping (RS) protocol, in which the replica makes a
random walk in a pre-defined set of temperatures $T^{1},...,T^{m}$.
Every so many MD/MC simulation steps, the replica jumps up or down
in temperature with equal probability. The temperature move is always
accepted unlike in ST. In this way temperature and configuration space
can be efficiently sampled with very widely spaced replicas providing
a good set of input trajectories for xTRAM.

Because there is no Metropolis-Hastings acceptance
criterion involved, the initial samples after each temperature swap
are definitely out of global, but also out of local equilibrium at
the new temperature. While discarding an initial fragment of the data
would seem to be a viable option, it turns out that instead using larger
lag times $\tau$ appears to work much better in correcting the estimates, as established for Markov model construction
\cite{Prinz2011}.
However, a solid theory for this observation
has yet to be found, which is beyond the scope of the current paper.

\section{Results}

To demonstrate the validity and resulting advantage of the
proposed estimator, two Langevin processes in model potentials, and
two explicitly solvated molecular dynamics processes are considered. In all cases we will compare three different estimators, which are the newly proposed xTRAM estimator, MBAR and histogram counting (direct counting estimate), each applied to the same sets of data.
Both accuracy and precision of all methods will be looked at by evaluating the systematic and statistical error
for representative discrete states and temperatures of interest.

\subsection{Two-well potential with solvent degrees of freedom}

\begin{figure*}[ht!]
\includegraphics[width=2\columnwidth]{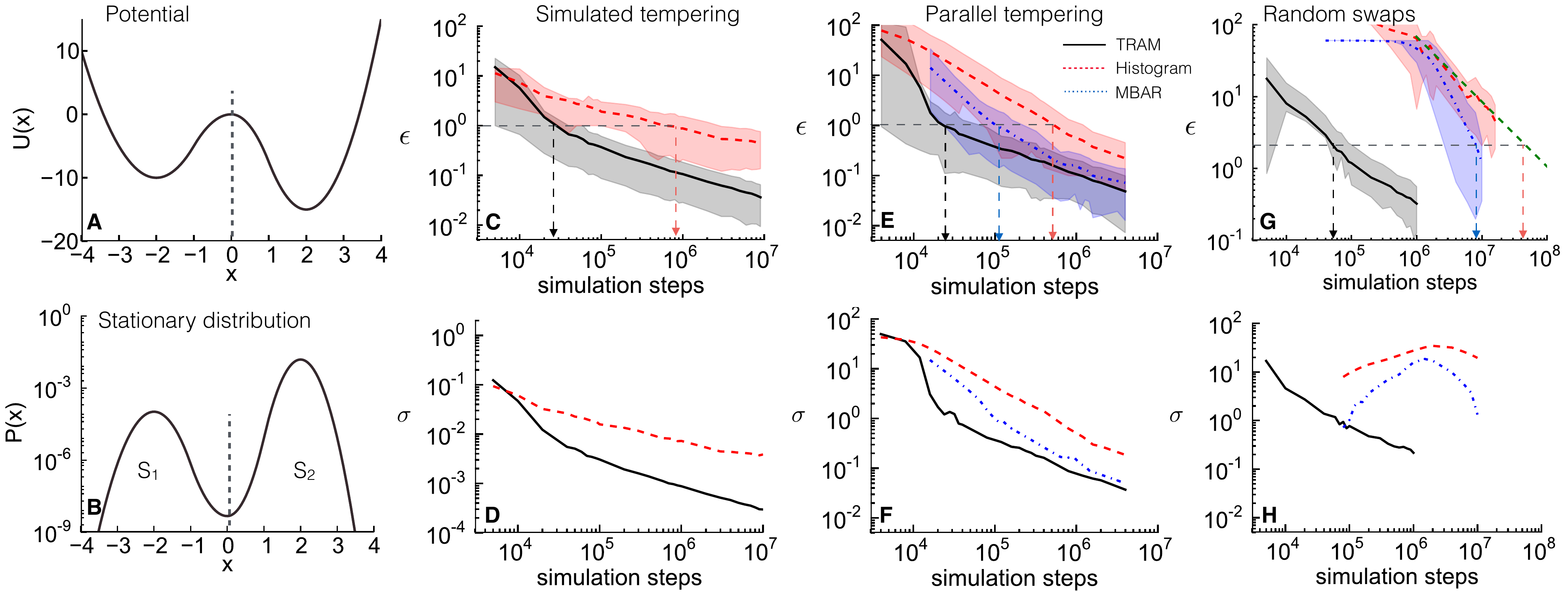}
\caption{(A) Double-well potential $U(x)$. (B) Corresponding stationary distribution at
a reduced temperature $k_B T = 1$. (C+D) Results of the ST simulation. (C) The
relative error of $\pi_{1}$ is shown w.r.t the number of simulation
steps for simulations with two solvent particles and four temperatures.
The direct histogram estimate (red, dashed) and xTRAM (black, continuous
line) are evaluated on the same data set. (D) Standard deviation ($\sigma$)
from 100 realizations are shown w.r.t. simulation steps taken for
direct count histogram (red, dashed) and xTRAM (black, continuous).
(E+F) Results of the PT simulation. (E) relative error $\pi_{1}$
w.r.t. to simulation steps from PT simulations over 100 realizations
for direct counts (red, dashed), MBAR (blue, dashed dotted)
and xTRAM (black continuous) (F) Standard deviation from data in (E).
(G+H) Results of a system with $N=50$ solvent particles using PT simulations for the MBAR and direct counting estimate and RS simulations for the xTRAM estimate. (G) relative error w.r.t. total number of simulation
steps for 20 temperature replica PT simulations direct count histograms (red,
dashed) MBAR (blue, dashed dotted) comparing to xTRAM (black, continuous)
analysis of RS simulation with six temperatures. The dashed green line
shows a fit of $t^{-0.5}$ to the tail of the relative error. (H) shows standard deviations of data in
(G).}
\label{fig:fig1} 
\end{figure*}

As a first example we consider Langevin
dynamics in an asymmetric double well potential (Fig.~\ref{fig:fig1}(A)) with 
corresponding stationary (Boltzmann) distribution $P(x)$ shown in Fig.~\ref{fig:fig1}(B)
for the reduced temperature $k_{B}T=1$. 
In order to make the system more complex, we add a set of $N$ solvent particles. 
Each solvent coordinate $i$ is subject to a harmonic potential $U(y_{i})=y_{i}^{2}$, where $y_{i}$
is the particle's position.

The state space is discretized into two states, corresponding to the two potential basins. 
We aim to estimate the equilibrium distribution of these two
states, from a set of different multi-temperature simulation protocols in combination with any of the estimators considered (xTRAM, MBAR and direct counting). All simulations are initiated from a local stationary distribution
in state $S_1$ and the three different simulation protocols chosen are parallel tempering (PT), simulated
tempering (ST) and random swapping (RS) simulations. With each simulation protocol 100 independent
realizations were generated and their results are shown in Fig.~\ref{fig:fig1}. For all three simulation protocols a temperature space needs to be defined, consisting of four exponentially spaced temperatures between $k_{B}T$=1
and $k_{B}T=10$ in reduced units, for fig.~\ref{fig:fig1}(C)-(F)
and six exponentially spaced temperatures between $k_{B}T=1$ and $k_{B}T=15$
in reduced units for fig.~\ref{fig:fig1}(G)-(H). The temperatures
are chosen in such a way that barrier crossings at the lowest temperature
are very rare events. For more details on the simulation protocols and setup see the
supplementary information.

Fig.~\ref{fig:fig1}(C-D) and (E-F) show the results
of ST and PT simulations with two solvent particles respectively.
The results are displayed in form of a log-log plot of the relative
error of the estimate of $\pi_{1}$ and its convergence behavior
with increased simulation time. The relative error is given by: 
\begin{equation}
	\epsilon=\Big|\frac{\pi(S_{1})_{\mathrm{exact}}-\pi(S_{1})_{\mathrm{estimate}}}{\pi(S_{1})_{\mathrm{exact}}}\Big|\label{eq:relative_error}.
\end{equation}
The stationary distribution, at the lowest reduced temperature of
$k_BT=1$ is obtained using all three estimators: (1) direct counting, (2) MBAR and (3) xTRAM. 
Fig.~\ref{fig:fig1}(C), (E) and (G) report averages and confidence intervals of the time-dependent
relative errors computed from $100$ realizations of each simulation.
Fig.~\ref{fig:fig1}(D), (F) and (H) report standard deviations ($\sigma$)
of the simulation data from 100 independent realizations and their
time dependence. In panels (C) and (E) the tail of the mean error
for all three methods decays approximately equal to $b/\sqrt{t}$,
where $b$ is a constant related to the decorrelation time $t_{\mathrm{corr}}$
required to generate an uncorrelated configuration at the
temperature analyzed. Arrows in panel (C) and (E) indicate a relative
error of 1. In the case of the ST simulation xTRAM outperforms direct
counting by a factor of 40 and in (E) for the parallel tempering simulation
xTRAM has a gain of a factor of five over MBAR estimates and a factor
of 25 over direct counting estimates. This means that the xTRAM estimates
converge up to a 40 fold faster in comparison to direct counts and
at least five fold faster in comparison to MBAR, indicating that the
decorrelation time with xTRAM can be much shorter. Consequently, less
simulation time needs to be invested when the data is analyzed with
xTRAM. Secondly the, standard deviation as seen in (D) and (F) is
consistently lower for xTRAM, meaning that over independent realizations,
the accuracy of the estimate is better in comparison to MBAR and
direct counting. 

Additional efficiency can be gained when the simple
random swapping (RS) simulation protocol can be employed instead of
ST or PT simulations, because then the number of replicas can be reduced
such that TRAM gives good results, while ST or PT replicas would not
mix well in temperature space. In Fig.~\ref{fig:fig1}(G) the results
of a simulation with $50$ solvent particles are depicted. In order
to achieve a good mixing in a PT simulation, 20 temperatures exponentially
spaced in the range of 1 to 15 in reduced units need to be used, which
is compared to a 6 replica RS simulation (see supplementary information for more details). The lag time $\tau$ for the evaluation could actually be chosen as small as the saving interval of the simulation in this case, resulting in the same convergence as using larger lag times. Looking at a relative error of $\epsilon=2$, an extrapolation needs
to be made to compute how many simulation steps are needed
for the PT direct counting estimate. From the extrapolated convergence
behavior it is found to be around $1\times10^{8}$ simulation steps.
Despite the fact that the RS protocol itself is not in equilibrium,
the correct equilibrium probabilities can be recovered when used in
conjunction with xTRAM. In this case, a relative error $\epsilon=2$
is achieved at around $1\times10^{5}$ simulation steps, indicating
an efficiency gain of around three orders of magnitude for RS/TRAM
and more than two orders of magnitude over MBAR used with the same PT simulation data as for the direct counts. It should be stressed again, that for MBAR and direct counting only simulations sampling from a global equilibrium can be used, therefore these estimators are not suitable to be used in conjunction with a random swapping simulation. 
Fig.~\ref{fig:fig1}(H) shows the standard deviation of the relative
error from 100 realizations. Initially, the standard deviation is deceptively small
for direct counts and MBAR, because many of the 100 simulations have only seen state 1. 
The standard deviation increases, as the second state is discovered reaching a peak, from which onwards $\sigma$
decreases again.

\subsection{Simple protein folding model}

In order to illustrate the limitations of the method, we now discuss an example where xTRAM offers no significant advantage
over the established MBAR estimator.
We consider an idealized folding potential
with an energetically stabilized native state and an entropically
stabilized denatured state. The state space is spanned by a vector
in $5$ dimensions, such that $\mathbf{{x}\in}\mathbb{R}^{5}$ and
$r=\vert\mathbf{x}\vert$ is the distance from zero. The potential
is defined as: 
\begin{equation}
U(r)=\begin{cases}
-2.5(r-3)^{2} & \mathrm{if}\; r<3\\
0.5(r-3)^{3}-(r-3)^{2} & \mathrm{if}\; r\ge3\label{eq:folding_pot}
\end{cases}
\end{equation}
and depicted in Fig.~\ref{fig:fig2}(A). Again a Langevin dynamics
simulation was carried out with more details provided in the supplementary
information. The system is discretized into two states: native and
denatured, with a state boundary at $r=2.7$, representing the distance
around which the lowest probability density is observed. All simulations are initiated in the native state.

\begin{figure*}[ht!]
\includegraphics[width=2\columnwidth]{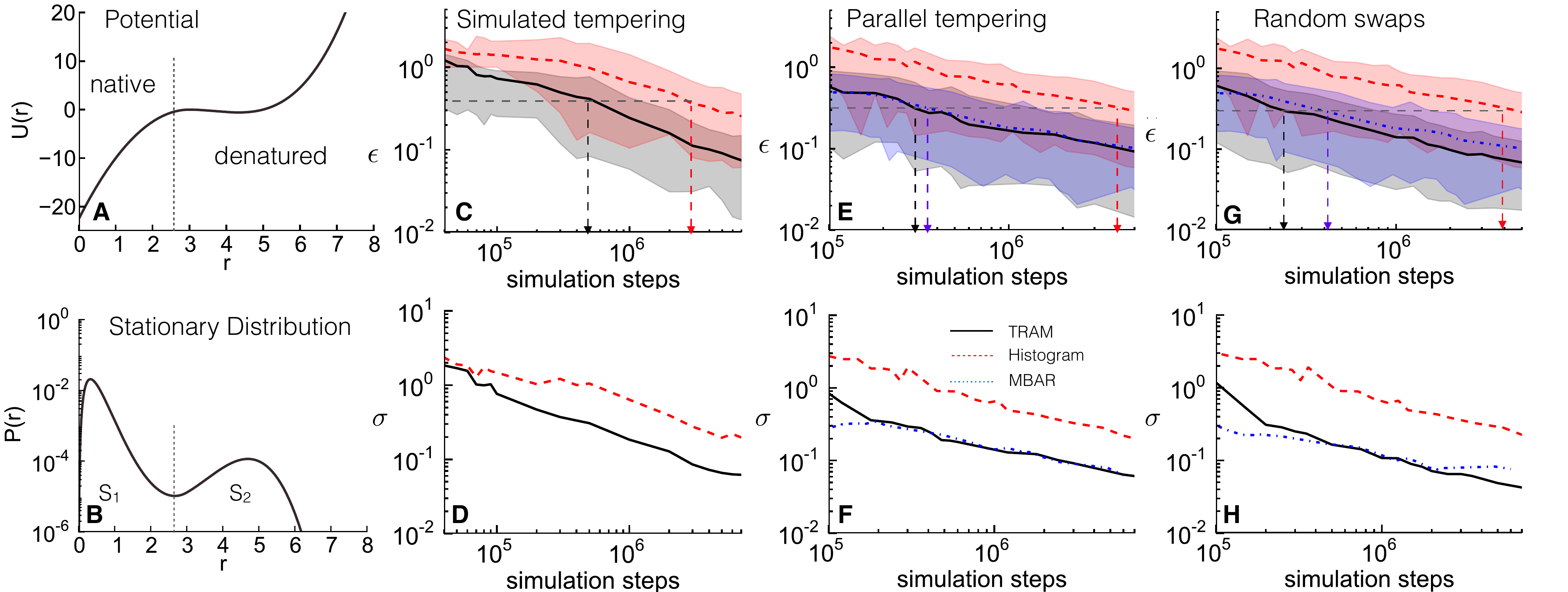}
\caption{(A) Potential $U(r)$. (B) Corresponding stationary distribution at
$\frac{1}{\beta}=1.1k_{B}T$. Results are shown in terms of the relative
error $\epsilon$ of the probability of being in the ``denatured''
state versus the number of simulation steps required. (C)-(D) Results
of the ST simulation. (C) black continuous line xTRAM estimate, red dashed line direct counting estimate. Arrows indicate an $\epsilon=0.3$ error level (D) standard deviation of the 100 realizations from (C) for both estimators. (E)-(F) Results of the parallel tempering simulation. (E) In blue dashed dotted the MBAR estimate, arrows indicate $\epsilon=0.2$ error level. (F) standard deviation of data in (E) from 100 realizations. (G)-(H) Comparison of RS + xTRAM estimate to the PT estimate using MBAR and direct counts in (E). Arrows indicate $\epsilon=0.2$ error level. (H) standard deviation of the data in (G) over 100 realizations. }
\label{fig:fig2}
\end{figure*}

The potential and the exact stationary density $\pi(r)$
at $\frac{1}{\beta}=1.1k_{B}T$ are shown in Fig.~\ref{fig:fig2}(A)
and (B) respectively. Note that the denatured state is stabilized
by entropy, as more microstates are available for $r>2.7$ than for
$r<2.7$. The convergence of the estimate of the relative error, Eq.~(\ref{eq:relative_error})
of the unfolded state is measured for a set of ST, PT and RS simulations.
Results are taken from $100$ different realizations and six exponentially
spaced temperatures between $\frac{1}{\beta}=1.1k_{B}T$ and $\frac{1}{\beta}=1.7k_{B}T$.
Panel (C) and (D) of Fig.~\ref{fig:fig2} show the results of the ST simulation, comparing the convergence of direct counting against xTRAM of the relative error of being in the denatured state. As before, xTRAM is shown by the black, continuous line, direct counting by the dashed, red line. Arrows indicating an error level of $\epsilon=0.3$ are used as guidance for the comparison of the convergence. Using xTRAM as the estimator for the analysis of the simulation results in a nine fold gain over direct counting. Shaded areas indicate confidence intervals. Fig.~\ref{fig:fig2}(D) shows the convergence of the standard deviation obtained from 100 independent realizations of the ST simulation from (C). The standard deviation of the xTRAM estimate is consistently lower than for the direct counting estimate. Fig.~\ref{fig:fig2}(E) and (F) summarize the results of the parallel tempering simulations. Here an 11 fold gain is observed when using xTRAM over direct counting, but MBAR and xTRAM perform almost equally as well, indicated by the arrows shown at an error level of $\epsilon=0.2$. This behavior suggests, that in this model samples from the local and global equilibrium are generated at the same rate. Fig.~\ref{fig:fig2} (F) shows the standard deviation of the data in (E) from the 100 independently generated simulations. 


Fig.~\ref{fig:fig2}(G) and (H), compare the direct counting and MBAR estimate of (E) with an RS simulation using only a single replica and 4 exponentially spaced temperatures between $k_BT=[1.1\ldots1.7]$. Now xTRAM has a 2 fold gain over MBAR. 

As xTRAM is a local-equilibrium generalization of MBAR, it is guaranteed to have equal or better estimation accuracy. However, the results above indicate
that the accuracy gain of xTRAM over MBAR can be small in some systems. In the folding potential this is presumably
due to the fact that the different temperatures not only help in barrier crossing, but give rise to vastly different equilibrium probabilities of the folded state
(stable at low temperatures) and the unfolded state (stable at high temperatures). Thus, even at short simulation times, both the folded and unfolded states 
are present in the replica ensemble and can successfully be reweighted using MBAR without relying on too many actual configuration state transitions.
xTRAM will gain efficiency in situations, where the state space exploration is slowed down by higher friction or additional barriers, as often found in
macromolecules.


\subsection{Alanine dipeptide}

In order to test the xTRAM estimator on a system
with many degrees of freedom, we turn to molecular dynamics (MD) simulations
using an explicit solvent model. To this end we study solvated alanine
dipeptide, a small and well-studied peptide with multiple metastable
states~\cite{Chekmarev2004,Smith1999,Chodera2006,Du2011} and around
$6000$ degrees of freedom in the case of the system set-up used here. Alanine dipeptide was prepared using an explicit water model and simulated in the MD software package OpenMM~\cite{OpenMM}.
All the necessary simulation details are provided in the appendix. 

\begin{figure}[h!]
\includegraphics[height=0.83\textheight]{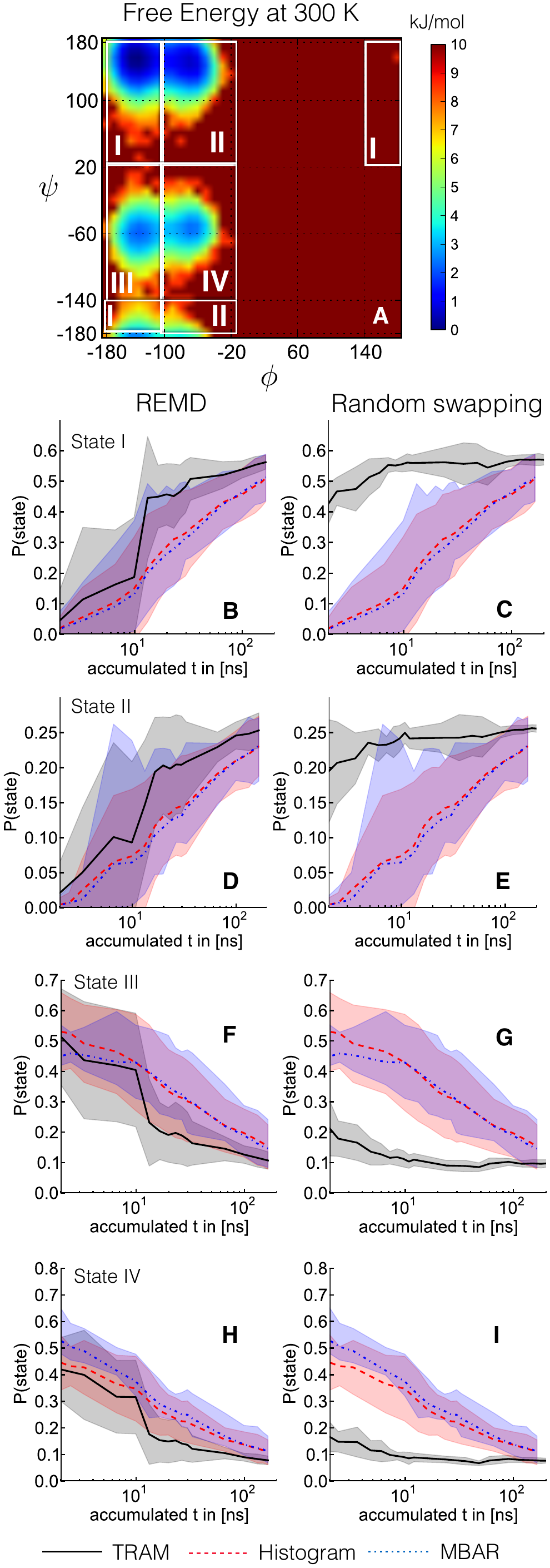}
\caption{(A) Free energy with a 10 $\mathrm{kJ/,mol}$ cutoff, indicating discrete states. (B, D, F, H) REMD simulation convergence of state probability over ten realizations. Estimates are obtained from xTRAM (black, continuous line), MBAR (blue, dashed-dotted line), and direct counting (red, dashed line). (C, E, G, I) RS simulation convergence of state probabilities over five realizations, compared to REMD simulation estimates using MBAR and direct counts.}
\label{fig:free_energy} 
\end{figure}
The dominant conformations of this system are the different rotamers set by the
dihedral angles $\psi$ and $\phi$. This means we are interested in estimating the free energy surface in $\phi$/$\psi$ space at a low temperature of interest. 
Again, we are interested in extracting the stationary
probabilities of metastable basins at different temperatures. However,
the system at $T=300\, K$ is not very metastable, thus we introduce
an artificial metastability to the torsional angles. For
this purpose we add a potential to the minima of the $\phi$
and $\psi$ dihedral angles in order to extend the time the system
stays in a particular angular configuration. The ideal choice of such
an additional potential are periodic von Mises potentials of the form:
\begin{equation}
U(\delta)=\epsilon\exp\Big(\frac{\kappa\cos(\delta-\delta_{0})}{2\pi I_{0}}\Big),
\end{equation}
where $I_{0}$ is a zeroth order Bessel function and $\delta$ is
the angle to which the additional potential is added. For more details
see the supplementary information. 
We use two different types of multi-temperature
simulations: a set 10 independent realizations of a 32 temperature
REMD simulation with temperatures exponentially spaced in the range
of $T=[300K\,-600K]$ and a set of independent independent realizations of a
13 temperature RS simulation, where only every third temperature out
of the REMD multi-temperature ensemble was used. 
From the 10 REMD trajectories the last 1 ns of each 
were used to estimate the free energy surface in dihedral
$\phi$ and $\psi$ space as shown in Fig.~\ref{fig:free_energy}(A).
From the free energy surface four discrete states could be defined,
numbered, and highlighted by the white boxes. All simulations were
initiated in state IV. From the simulations dihedral coordinates, discrete
trajectories were generated which then allow a stationary estimate
through direct counts of the frequency of each state visited along
the trajectory (in the case of the REMD simulation). The same discrete
trajectories are also used for xTRAM and MBAR estimation. 

For the RS simulation the total simulation time
was less, as only 13 instead of the 32 parallel replicas were simulated.
Confidence intervals are indicated by the shaded regions calculated
over the independent realizations of every simulation type. 
In order for the RS simulation to produce valid
results the lag-time at which the data points are used needs to be
adjusted, the saving interval of the trajectory was 0.1 ps and and
the lag interval at which data frames were used was chosen to be $\tau=$1
ps and temperature switches were carried out every 10 ps, for more
details on the RS simulation refer to the supplementary information. 

Fig.~\ref{fig:free_energy}(B), (D), (F), and (H)
show the convergence results of the REMD simulation. While all
estimators yield similar (and inaccurate) estimates for very short simulation times, 
xTRAM exhibits considerable advantages over MBAR and direct counting after 10 ns 
simulation time. The fastest-converging estimator (see below) produces stable 
equilbrium probabilities of about $(0.57, 0.25, 0.13, 0.1)$ for states 1-4 at 100 ns simulation 
time. Using REMD data, xTRAM converges to within about 10\% of these values with roughly
one order of magnitue simulation data compared to MBAR and direct counting 
(20 ns versus 200 ns).

Fig.~\ref{fig:free_energy}(C), (E), (G), and (I) compare the performance of the RS simulations when
analyzed with xTRAM, in comparison to the standard REMD simulations. As a result of
the smaller number of replicas required and the enhanced mixing properties, another
order of magnitude is gained with the RS protocol when comparing to the xTRAM estimate
of the REMD simulations. Since xTRAM is currently the only available estimator to unbias RS
simulations, the advantage of xTRAM over MBAR and direct counting amounts two orders
of magnitude (xTRAM with RS versus MBAR with REMD). This advantage of xTRAM in conjunction
with RS can be much larger for systems with many degrees of freedom, where a REMD simulation
would need many closely spaced replicas, thus resulting in vast computational effort and slow
exchange dynamics.

\subsection{Deca-alanine}

Finally we consider the 10 amino acid long deca-alanine (Ala$_{10}$). 
This peptide is know to undergo a helix-coil transition, which has been studied extensively~\cite{Best2009,Michaud-Agrawal2011,Sorin2005,Huang2001}.
Ten independent runs of all-atom replica exchange simulations were conducted with the GROMACS software MD package, each using 24 exponentially spaced temperatures ranging
from T = 290 K to 400 K~\cite{Hess2008}. We ran the simulation for 40 ns total simulation
time per replica and conducted 10 independent realizations of these. For more detailed description
of the simulation details see the supplementary information. 

\begin{figure}[hT!]
\includegraphics[width=1\columnwidth]{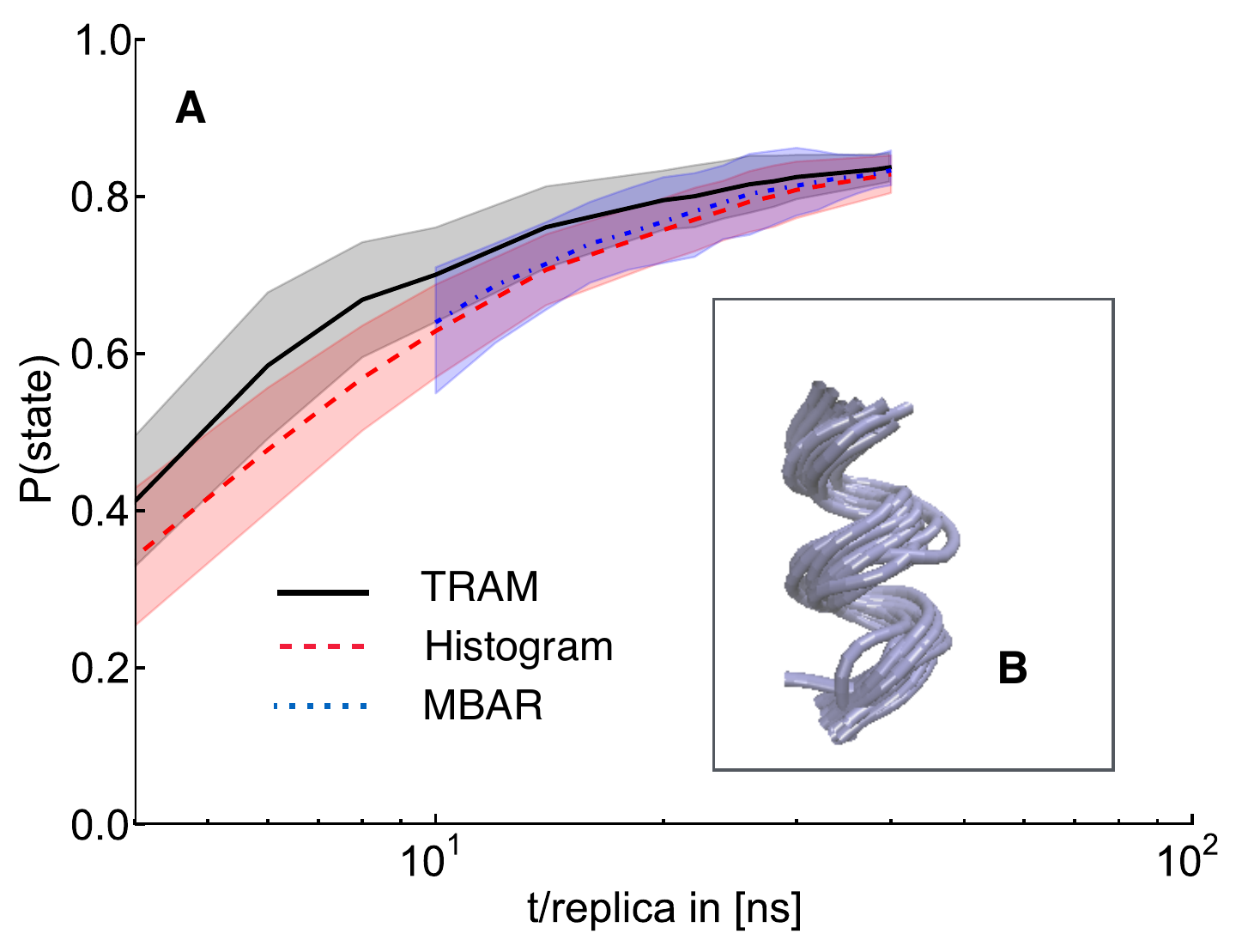}
\caption{(A) Convergence of the probability of the system being in a helical state with respect to the total simulation time per replica. The xTRAM estimate is given by the continuous black line, the histogram count by the dashed red line, and the MBAR estimate by the dashed-dotted blue line. Confidence intervals are obtained from ten independent REMD simulations. (B) Structure averages of the helix state whose convergence is shown in (A).}
\textcolor{black}{\label{fig:deca-alanine} }
\end{figure}

In this larger molecular system the discretization of configuration space is no longer trivial. 
For this purpose we use time-lagged independent component analysis (TICA) on the replica trajectories of the set of C$_{\alpha}$ distances, 
omitting nearest neighbor distances along the peptide chain~\cite{Perez-Hernandez2013}. TICA is useful to identify the subspace in which
the slowest transitions occur. Here, we chose three leading TICA coordinates,
and used a regular spatial clustering on these with a minimum distance
cutoff of 3 yielding 44 discrete clusters. This analysis was carried using the Markov model package EMMA~\cite{Senne2012}.
See supplementary information for more details. 

xTRAM, MBAR and histogram counting were used in order to estimate the equilibrium probabilities on the 44 discrete configuration states. 
In order to obtain a simple representation of the results, the equilibrium probabilities summed over all $\alpha$-helical states
is shown in Fig. \ref{fig:deca-alanine}(B). As before, we are interested in the analysis at the lowest simulation temperature ($T$ = 290 K).

As seen in Fig~\ref{fig:deca-alanine}(A) the advantage gained from TRAM in comparison to MBAR and direct counting in this case is only
about two fold. However, this can be attributed to the fact, that the
system does not display a very strong metastability and the slowest
timescale, computed independently with a Markov state model on direct 290 K trajectories, 
is only 14 ns. 
Moreover, like the simple folding model above, Ala$_{10}$ has the same property that the temperatures stabilize the folded and
unfolded states quite differently, leading to simultaneous observation of folded and unfolded states at early stages of the replica
simulation, and also to the fact that significantly many transitions between folded and unfolded state occur only in a very small part of the replica
ensemble (in the replicas around the melting point). As a result, the advantage of taking configuration-state transitions into account is smaller in this case compared to systems at which the different metastable states are present at a larger range of temperatures.
As demonstrated for the simple folding model above, larger gains of computational efficiency can still be obtained by reducing the number of replicas, 
e.g. by employing the RS protocol in conjunction with the xTRAM estimator.


\section{Discussion and Conclusion}

Expanded TRAM can be used to obtain estimates of
equilibrium properties from any set of simulations that was conducted
at different thermodynamic states, such as multiple temperatures,
Hamiltonians, or bias potentials, which we demonstrated here for multiple
temperature simulations. It is therefore applicable to generalized
ensemble simulations such as replica-exchange and parallel/simulated
tempering, as well as umbrella sampling or metadynamics. The quantities
estimated can include free energy differences, equilibrium probabilities
or equilibrium expectations of functions of configuration state. As
such, xTRAM has the same application scope as existing reweighting
estimators (e.g. WHAM and MBAR).

In contrast to WHAM and MBAR, xTRAM does not assume
simulation data to be generated from global equilibrium. Rather, xTRAM
combines ideas from MBAR and Markov modeling to an estimator that
makes use of both, Boltzmann reweighting of sampled configurations
between thermodynamic states, and transition count statistics between
different configuration states generated by contiguous trajectory
segments. Compared to MBAR, estimates obtained from xTRAM can be more
accurate as they suffer less from data that has not yet decorrelated
from the initial configurations, as well as more precise as the statistics
in the simulation data can be used more efficiently when every conditional
independent transition count is useful rather than only every globally
independent count.

xTRAM is an asymptotically correct estimator, i.e.
its estimates converge to the exact values in the limit of long or
many simulation trajectories. We have also shown that in the special
case when simulation data \emph{are}
at global equilibrium, we can derive the MBAR equations from the expectation
values of the xTRAM equations. MBAR is thus a special case of xTRAM,
suggesting that the accuracy of xTRAM estimates should be at least
equally good as those of MBAR estimates, but may be significantly better
when parts of the simulation data are not in global equilibrium.
The applications shown here confirm this result, exhibiting sometimes
order-of-magnitude more accurate estimates when xTRAM is employed,
and therefore allowing the use of shorter simulation times while maintaining the same accuracy in the estimate. 

While MBAR provides statistically optimal (i.e.
minimum-variance) estimates under the condition that data are in global
equilibria, it is currently not known whether xTRAM is also statistically
optimal. However, the applications in this paper suggest that the
variances of xTRAM estimates are in most cases significantly better than those of MBAR or direct counting.

An interesting aspect of xTRAM is the fact
that it does not rely on the data being at global equilibrium, thus
opening the door to consider new simulation methods that deliberately
sacrifice global equilibrium for enhanced sampling. 
This feature reflects the Markov-model nature of
the configuration-state statistics in xTRAM - Markov models also allow
to obtain unbiased estimates from short trajectories that are not
sampling from global equilibrium, as long as they sample from local
equilibrium within each configuration state. We have demonstrated
this ability by using xTRAM to unbias simple random swapping simulations
that exchange temperature replicas in complete ignorance of the Metropolis
acceptance probability. The hope is that with such a setup, large
systems can be simulated with very few widely spaced replicas, that
would be inappropriate for a PT or ST simulation that need energy
overlap between adjacent replicas. It was shown that with a sufficiently
large lag-time $\tau$ for evaluating the transition counts, xTRAM
provided accurate estimates with such a protocol, while achieving
a sampling efficiency that is orders of magnitude more efficient than
classical replica-exchange or parallel tempering simulations. In the
future, it will be necessary to develop a theory that quantifies the
local equilibrium error made by brute-force sampling protocols such
as random swapping in order to put their use on solid ground.

An implementation of xTRAM is available in pytram python package:
https://github.com/markovmodel/pytram

\begin{acknowledgments} We gratefully acknowledge
funding by SFB 958 and 1114 of the German Science Foundation, as well
as ERC starting grant pcCell of the European commission. We are indebted
to John D. Chodera for the help in setting up OpenMM simulations,
as well as Fabian Paul, Benjamin Trendelkamp-Schroer, Edina Rosta and John D. Chodera for useful
discussions. A. Mey is grateful for access to the University of
Nottingham High Performance Computing Facility. \end{acknowledgments}
\bibliography{MEEE}
\appendix
\onecolumngrid

\section{Proofs}

Here we proof the asymptotic convergence of xTRAM, and that the TRAM estimation
equations (25,26,27) of the main manuscript become identical to the MBAR equations for
the special case that each simulation samples from the global equilibrium
at its respective thermodynamic state.

\subsection{Asymptotic convergence of xTRAM}

In the statistical limit $N\rightarrow\infty$, we can use that the transition counts converge to their
conditional expectation values:
\begin{align}
\frac{c_{ij}^{I}}{N} & = \frac{N_{i}^{I}p_{ij}^{I}}{N}\\
\frac{b_{ij}^{I}}{N} & = \frac{N_{i}^{I}p_{i}^{IJ}}{N}.
\end{align}
Inserting these into the xTRAM estimator for equilibrium probabilities
results in the update:

\begin{align}
x_{i}^{I} & =\frac{1}{2}w^{I}\sum_{j=1}^{n}\frac{N_{i}^{I}p_{ij}^{I}+N_{j}^{I}p_{ji}^{I}}{\frac{N_{i}^{I}}{\pi_{i}^{I}}+\frac{N_{j}^{I}}{\pi_{j}^{I}}}+\frac{1}{2}\sum_{J=1}^{m}\frac{N_{i}^{I}p_{i}^{IJ}+N_{i}^{J}p_{i}^{JI}}{\frac{N_{i}^{I}}{w^{I}\pi_{i}^{I}}+\frac{N_{i}^{J}}{w^{J}\pi_{i}^{J}}},
\end{align}
using reversibility ($\pi_{i}^{I}p_{ij}^{I}=\pi_{j}^{I}p_{ji}^{I}$
and $w^{I}\pi_{i}^{I}p_{i}^{IJ}=w^{J}\pi_{i}^{J}p_{i}^{JI}$), we
get:
\begin{align}
x_{i}^{I} & =\frac{1}{2}w^{I}\sum_{j=1}^{n}\frac{N_{i}^{I}p_{ij}^{I}+N_{j}^{I}p_{ij}^{I}\frac{\pi_{i}^{I}}{\pi_{j}^{I}}}{\frac{N_{i}^{I}}{\pi_{i}^{I}}+\frac{N_{j}^{I}}{\pi_{j}^{I}}}+\frac{1}{2}\sum_{J=1}^{m}\frac{N_{i}^{I}p_{i}^{IJ}+N_{i}^{J}p_{i}^{IJ}\frac{w^{I}\pi_{i}^{I}}{w^{J}\pi_{i}^{J}}}{\frac{N_{i}^{I}}{w^{I}\pi_{i}^{I}}+\frac{N_{i}^{J}}{w^{J}\pi_{i}^{J}}}\\
 & =\frac{1}{2}w^{I}\pi_{i}^{I}\sum_{j=1}^{n}p_{ij}^{I}+\frac{1}{2}w^{I}\pi_{i}^{I}\sum_{J=1}^{m}p_{i}^{IJ}\\
 & =w^{I}\pi_{i}^{I}.
\end{align}

\subsection{Free energies}

In order to show the relationship between TRAM and MBAR, we use the
TRAM equations and specialize them using the MBAR assumption that
each thermodynamic state is sampled from global equilibrium. In order
to relate the TRAM and MBAR free energy estimates, $f^{I}$, we merge
all configuration states to one state. When converged, the TRAM equations
(25,26,27) of the main manuscrip then become:
\begin{align}
\pi^{I}p^{IJ} & =\frac{b^{IJ}+b^{JI}}{\frac{N^{I}}{\pi^{I}}+\frac{N^{J}}{\pi^{J}}}\\
0 & =-\ln\frac{N}{N^{I}}\pi^{I}.
\end{align}
From the second equation, we obtain $\pi^{I}=N^{I}/N$. Inserting
into the first equation yields:
\begin{equation}
2 N^{I} p^{IJ} = b^{IJ}+b^{JI}.
\end{equation}
summing over $J$ on both sides:
\begin{align}
2N^{I} & =\sum_{J}(b^{IJ}+b^{JI})\\
 & =N^{I}+\sum_{J}b^{JI}\\
1 & =\sum_{J}\frac{b^{JI}}{N^{I}}.
\end{align}
Inserting the TRAM definition for temperature transition counts
\begin{equation}
b^{JI}=\sum_{\mathbf{x} \in S^J}\frac{N^{I}e^{f^{I}-u^{I}(\mathbf{x})}}{\sum_{K}N^{K}e^{f^{K}-u^{K}(\mathbf{x})}}
\end{equation}
results in:
\begin{align}
1 & =\sum_{J}\sum_{x \in S^J}\frac{e^{f^{I}-u^{I}(\mathbf{x})}}{\sum_{K}N^{K}e^{f^{K}-u^{K}(\mathbf{x})}},
\end{align}
and thus:
\begin{align}
e^{-f^{I}} & =\sum_{\mathrm{all}\:\mathbf{x}}\frac{e^{-u^{I}(\mathbf{x})}}{\sum_{K}N^{K}e^{f^{K}-u^{K}(\mathbf{x})}}\label{eq:MBAR_partition_function-1}\\
f^{I} & =-\ln\sum_{\mathrm{all}\:\mathbf{x}}\frac{e^{-u^{I}(\mathbf{x})}}{\sum_{K}N^{K}e^{f^{K}-u^{K}(\mathbf{x})}}\label{eq:MBAR_free_energy}
\end{align}
 which is exactly the MBAR estimator for the reduced free energy of
thermodynamic state $I$ (See Eq. (11) in \cite{Shirts2008}).

\subsection{Temperature-state transitions and resulting state probability expectations}

We want to derive an expression for the normalized stationary probabilities
$\pi_{i}^{I}$ that results from xTRAM under the assumption of sampling
from global equilibrium within each of the thermodynamic states $I$
and compare this to the corresponding xTRAM expectation. 

In order to find the xTRAM estimations of $\pi_{i}^{I}$, we write
down the reversible transition matrix optimality conditions. For transitions
between thermodynamic states we have, using that the absolute stationary
probability vector is given by elements $N^{I}\pi_{i}^{I}/N$:

\begin{eqnarray}
N^{I}\pi_{i}^{I}p_{i}^{IJ} & = & \frac{b_{i}^{IJ}+b_{i}^{JI}}{\frac{N_{i}^{I}}{N^{I}\pi_{i}^{I}}+\frac{N_{i}^{J}}{N^{J}\pi_{i}^{J}}}\\
N^{I}\pi_{i}^{I}p_{i}^{IJ} & = & N^{I}\pi_{i}^{I}N^{J}\pi_{i}^{J}\frac{b_{i}^{IJ}+b_{i}^{JI}}{N^{J}\pi_{i}^{J}N_{i}^{I}+N^{I}\pi_{i}^{I}N_{i}^{J}}\\
p_{i}^{IJ} & = & N^{J}\pi_{i}^{J}\frac{b_{i}^{IJ}+b_{i}^{JI}}{N^{J}\pi_{i}^{J}N_{i}^{I}+N^{I}\pi_{i}^{I}N_{i}^{J}}.
\end{eqnarray}
Using global equilibrium and the statistical limit $N\rightarrow\infty$
allows us to write: $N_{i}^{I}=\pi_{i}^{I}N^{I}$, $N_{i}^{J}=\pi_{i}^{J}N^{J}$,
and $b_{i}^{IJ}=N_{i}^{I}p_{i}^{IJ}$. Inserting these equations yields:
\begin{align}
\frac{b_{i}^{IJ}}{N_{i}^{I}} & =N^{J}\pi_{i}^{J}\frac{b_{i}^{IJ}+b_{i}^{JI}}{N^{J}\pi_{i}^{J}\pi_{i}^{I}N^{I}+N^{I}\pi_{i}^{I}\pi_{i}^{J}N^{J}}\\
 & =\frac{b_{i}^{IJ}+b_{i}^{JI}}{2\pi_{i}^{I}N^{I}}\\
b_{i}^{IJ} & =\frac{b_{i}^{IJ}+b_{i}^{JI}}{2}\\
b_{i}^{JI} & =b_{i}^{IJ}.
\end{align}
Using the TRAM estimators for thermodynamic-state transition counts:
\begin{eqnarray}
\hat{b}_{i}^{IJ} & = & \sum_{\mathbf{x}\in S_{i}^{I}}\frac{N^{J}e^{f^{J}-u^{J}(\mathbf{x})}}{\sum_{K}N^{K}e^{f^{K}-u^{K}(\mathbf{x})}}\\
\hat{b}_{i}^{JI} & = & \sum_{\mathbf{x}\in S_{i}^{J}}\frac{N^{I}e^{f^{I}-u^{I}(\mathbf{x})}}{\sum_{K}N^{K}e^{f^{K}-u^{K}(\mathbf{x})}},
\end{eqnarray}
we have the equality:
\begin{equation}
\sum_{\mathbf{x}\in S_{i}^{I}}\frac{N^{J}e^{f^{J}-u^{J}(\mathbf{x})}}{\sum_{K}N^{K}e^{f^{K}-u^{K}(\mathbf{x})}}=\sum_{\mathbf{x}\in S_{i}^{J}}\frac{N^{I}e^{f^{I}-u^{I}(\mathbf{x})}}{\sum_{K}N^{K}e^{f^{K}-u^{K}(\mathbf{x})}}.
\end{equation}
Summing over $J$, it follows that
\begin{eqnarray}
\sum_{\mathbf{x}\in S_{i}^{I}}\frac{\sum_{J}N^{J}e^{f^{J}-u^{J}(x)}}{\sum_{K}N^{K}e^{f^{K}-u^{K}(x)}} & = & \sum_{J}\sum_{\mathbf{x}\in S_{i}^{J}}\frac{N^{I}e^{f^{I}-u^{I}(\mathbf{x})}}{\sum_{K}N^{K}e^{f^{K}-u^{K}(\mathbf{x})}}\\
N_{i}^{I} & = & \sum_{J}\sum_{\mathbf{x}\in S_{i}^{J}}\frac{N^{I}e^{f^{I}-u^{I}(\mathbf{x})}}{\sum_{K}N^{K}e^{f^{K}-u^{K}(\mathbf{x})}}\\
\frac{N_{i}^{I}}{N_{I}}e^{-f^{I}} & = & \sum_{J}\sum_{\mathbf{x}\in S_{i}^{J}}\frac{e^{-u^{I}(\mathbf{x})}}{\sum_{K}N^{K}e^{f^{K}-u^{K}(\mathbf{x})}}.
\end{eqnarray}
Using again $N_{i}^{I}=\pi_{i}^{I}N^{I}$ we rewrite this equation
into:
\begin{equation}
\pi_{i}^{I}=\frac{\sum_{J}\sum_{\mathbf{x}\in S_{i}^{J}}\frac{e^{-u^{I}(\mathbf{x})}}{\sum_{K}N^{K}e^{f^{K}-u^{K}(\mathbf{x})}}}{e^{-f^{I}}},
\end{equation}
using Eq. (\ref{eq:MBAR_partition_function-1}) this results in 
\begin{equation}
\pi_{i}^{I}=\frac{\sum_{\mathbf{x}\in S_{i}}\frac{e^{-u^{I}(\mathbf{x})}}{\sum_{K}N^{K}e^{f^{K}-u^{K}(\mathbf{x})}}}{\sum_{\mathbf{x}\in\Omega}\frac{e^{-u^{I}(\mathbf{x})}}{\sum_{K}N^{K}e^{f^{K}-u^{K}(\mathbf{x})}}}
\end{equation}
which is exactly the MBAR expectation value (using Eqs. (14-15) in
\cite{Shirts2008} with the indicator function of set
$S_{i}$ as function $A(\mathbf{x})$).

\subsection{MBAR expectation values from TRAM}

Given the local free energies $f_{i}^{I}=f^{I}-\ln\pi_{i}^{I}$ computed
from xTRAM, we consider each combination of configuration state and
thermodynamic state as a thermodynamic state for MBAR and use the
MBAR equations to compute expectation values \cite{Shirts2008}.
We define the weights:

\begin{equation}
g(\mathbf{x})=\frac{\mathrm{e}^{-u(\mathbf{x})}}{\sum_{K}\sum_{i}N_{i}^{K}\mathrm{e}^{f_{i}^{K}-u^{K}(\mathbf{x})}},
\end{equation}
and then obtain expectation values of the function $A(x)$ as:
\begin{equation}
\mathbb{E}[A]=\frac{\sum_{\mathbf{x}}g(\mathbf{x})\, A(\mathbf{x})}{\sum_{\mathbf{x}}g(\mathbf{x})}.\label{eq:mbar_expectation}
\end{equation}
This choice can be motivated as follows: Using $f_{i}^{K}=f^{K}-\ln\pi_{i}^{K}$
we obtain
\begin{align}
g(\mathbf{x}) & =\frac{\mathrm{e}^{-u(\mathbf{x})}}{\sum_{K}\sum_{i}N_{i}^{K}\mathrm{e}^{f^{K}-\ln\pi_{i}^{K}-u^{K}(\mathbf{x})}}\\
 & =\frac{\mathrm{e}^{-u(\mathbf{x})}}{\sum_{K}\sum_{i}\frac{N_{i}^{K}}{\pi_{i}^{K}}\mathrm{e}^{f^{K}-u^{K}(\mathbf{x})}}.
\end{align}
We now choose $N_{i}^{K}=\pi_{i}^{K}N^{K}$ (statistical limit and
global equilibrium) and obtain:
\begin{equation}
g(\mathbf{x})=\frac{\mathrm{e}^{-u(\mathbf{x})}}{n\sum_{K}N^{K}\mathrm{e}^{f^{K}-u^{K}(\mathbf{x})}},
\end{equation}
where the factor $n^{-1}$ cancels in Eq. (\ref{eq:mbar_expectation}).
We thus get exactly the MBAR equation for an expectation value (compare
to Eqs. (14-15) in \cite{Shirts2008}).

\section{Estimation of free energies}

\subsection{Initial choice for the free energies $f^{I}$}

We first seek a way to come up with an initial guess of the free energies
$f^{I}$ for all thermodynamic states $I$. Here we follow the approach
of Bennett \cite{Bennett_JCompPhys76_BAR}. We first order the different
thermodynamic states simulated at in a sequence $(1,\:...,I,\: I+1,\:...,\: m)$,
e.g. ascending temperatures, and then construct a reversible Metropolis-Hastings
Monte Carlo process between neighboring thermodynamic states. We define
the Metropolis function $M(x)=\min\{1,\exp(-x)\}$ and request the
detailed balance equation to be fulfilled: 
\begin{equation}
M\left(u^{I+1}(\mathbf{x})-u^{I}(\mathbf{x})\right)\mathrm{e}^{-u^{I}(\mathbf{x})}=M\left(u^{I}(\mathbf{x})-u^{I+1}(\mathbf{x})\right)\mathrm{e}^{-u^{I+1}(\mathbf{x})}.
\end{equation}
Integrating over the configuration space $\Omega$ and multiplying
the left-hand side by $Z^{I}/Z^{I}$ and the right-hand side by $Z^{I+1}/Z^{I+1}$
yields
\begin{equation}
Z^{I}\frac{\int d\mathbf{x}\: M\left(u^{I+1}(\mathbf{x})-u^{I}(\mathbf{x})\right)\mathrm{e}^{-u^{I}(\mathbf{x})}}{Z^{I}}=Z^{I+1}\frac{\int d\mathbf{x}\: M\left(u^{I}(\mathbf{x})-u^{I+1}(\mathbf{x})\right)\mathrm{e}^{-u^{I+1}(\mathbf{x})}}{Z^{I+1}},
\end{equation}
and thus we can derive an estimator for the ratio of neighboring partition
functions 
\begin{equation}
\frac{\hat{Z}^{I+1}}{\hat{Z}^{I}}=\frac{\left\langle M\left(u^{I+1}(\mathbf{x})-u^{I}(\mathbf{x})\right)\right\rangle _{I}}{\left\langle M\left(u^{I}(\mathbf{x})-u^{I+1}(\mathbf{x})\right)\right\rangle _{I+1}},\label{eq:estimator_F0}
\end{equation}
where $\left\langle \cdot\right\rangle _{I}$ denotes the expectation
value at thermodynamic state $I$. Using $f^{I}=-\ln Z^{I}$, we get:
\begin{equation}
f^{I+1}=f^{I}-\ln\frac{\left\langle M\left(u^{I+1}(\mathbf{x})-u^{I}(\mathbf{x})\right)\right\rangle _{I}}{\left\langle M\left(u^{I}(\mathbf{x})-u^{I+1}(\mathbf{x})\right)\right\rangle _{I+1}}.\label{eq:f_I+1_from_f_I}
\end{equation}
All $f^{I}$ can be shifted by an arbitrary additive constant. We
thus set $f^{1}=0$ and compute all $f^{2},...,f^{m}$ by iterative
application of Eq. (\ref{eq:f_I+1_from_f_I}).

\subsection{Corrections to the free energy}

Suppose in the previous ($k$th) iteration, we were using the estimate
\begin{equation}
(f^{1\,(k)},...,f^{m\,(k)})\Leftrightarrow(Z^{1\,(k)},...,Z^{m\,(k)})
\end{equation}
and after evaluation of transition counts and estimation of the transition
matrix, our estimated stationary distribution vectors $\pi_{i}^{I}$
sum up to:
\begin{eqnarray}
\sum_{i}({\tilde{\pi}}_{i}^{I})^{\:(k)}  & = & \frac{w^{I}}{\sum_{K}w^{K}}\frac{1}{Z^{I\,(k)}}\int_{\Omega}d\mathbf{x}\: e^{-u^{I}(\mathbf{x})}\\
 & = & \frac{w^{I}}{\sum_{K}w^{K}}\frac{Z^{I}}{Z^{I\,(k)}},
\end{eqnarray}
which suggests the update rules
\begin{eqnarray}
Z^{I\,(k+1)} & = & Z^{I\,(k)}\frac{\sum_{K}w^{K}}{w^{I}}\sum_{i}\left(\tilde{\pi}_{i}^{I}\right)^{(k)},
\end{eqnarray}
and using $f^{I}=-\ln Z^{I}$ we get
\begin{eqnarray}
-\ln Z^{I\,(k+1)} & = & -\ln Z^{I\,(k)}-\ln\frac{\sum_{K}w^{K}}{w^{I}}\sum_{i}\left(\tilde{\pi}_{i}^{I}\right)^{(k)}\\
f^{I\,(k+1)} & = & f^{I\,(k)}-\ln\frac{\sum_{K}w^{K}}{w^{I}}\sum_{i}\left(\tilde{\pi}_{i}^{I}\right)^{(k)}.
\end{eqnarray}

\section{Computation of equilibrium probabilities}

Suppose we are given the count matrix 
\begin{equation}
\mathbf{\tilde{N}}=\left(\begin{array}{cccc}
\mathbf{C}^{1}+\mathbf{B}^{1,1} & \mathbf{B}^{1,2} & \cdots & \mathbf{B}^{1,m}\\
\mathbf{B}^{2,1} & \mathbf{C}^{2}+\mathbf{B}^{2,2} & \ddots & \vdots\\
\vdots & \ddots & \ddots & \mathbf{B}^{m-1,m}\\
\mathbf{B}^{m,1} & \cdots & \mathbf{B}^{m,m-1} & \mathbf{C}^{m}+\mathbf{B}^{m,m}
\end{array}\right),
\end{equation}
containing the diagonal blocks with configuration-state transition
counts $\mathbf{C}^{I}=(c_{ij}^{I})$, $i,j=1,...,n$, and the off-diagonal
blocks containing diagonal matrices with thermodynamic-state transition
counts $\mathbf{B}^{I,J}=\mathrm{diag}(b_{i}^{IJ})$, $i=1,...,n$.
We here assume that these expanded dynamics in configuration and thermodynamic
state space are reversible, and thus estimate a transition matrix
$\mathbf{\tilde{P}}$ with maximum likelihood given $\mathbf{\tilde{N}}$
under the detailed balance constraints. Subsequently the stationary
distribution $\boldsymbol{\tilde{\pi}}$ is computed, which is the
quantity of interest. Due to the detailed balance constraints, we
cannot give a closed expression for $\mathbf{\tilde{P}}$. However,
an efficient iterative estimator that computes both $\mathbf{\tilde{P}}$
and $\boldsymbol{\tilde{\pi}}$ is described in \cite{Prinz2011}.

Since we are primarily interested in estimating $\boldsymbol{\tilde{\pi}}$
and not $\mathbf{\tilde{P}}$, we here describe a simple direct estimator
for $\boldsymbol{\tilde{\pi}}$. In general, given a count matrix
$\mathbf{N}\in\mathbb{R}^{r\times r}$ and $r$ in this special case representing the spatial and thermodynamic extension of the countmatrix, i.e. $r=n\times m$, let $\mathbf{P}$ be the associated maximum likelihood
reversible transition matrix and $\boldsymbol{\pi}>0$ its stationary
distribution. Defining $x_{ij}=\alpha\pi_{i}p_{ij}$ with an arbitrary constant $\alpha > 0$, 
the following equations are fulfilled at the optimum:
\begin{equation}
x_{ij}=\frac{n_{ij}+n_{ji}}{\frac{n_{i}}{\sum_{k=1}^{r}x_{ik}}+\frac{n_{j}}{\sum_{k=1}^{r}x_{jk}}}
\end{equation}
for all $i,j=1,...,r$ and for all $\alpha>0$. Realizing that $\sum_{k}x_{ik}=\alpha\pi_{i}$
and summing both sides over $j$, we obtain:
\begin{equation}
\pi_{i}=\sum_{j=1}^{r}\frac{n_{ij}+n_{ji}}{\frac{n_{i}}{\pi_{i}}+\frac{n_{j}}{\pi_{j}}}.
\end{equation}
In order to arrive at an estimator for $\pi_{i}$, we define the variables
$x_{i}$ that are supposed to converge towards $\pi_{i}$ and write
down the fixed point iteration:
\begin{align}
x_{i}^{(k)} & =\sum_{j=1}^{r}\frac{n_{ij}+n_{ji}}{\frac{n_{i}}{\pi_{i}^{(k)}}+\frac{n_{j}}{\pi_{j}^{(k)}}}\label{eq:pi_gen_iteration_1}\\
\pi_{i}^{(k+1)} & =\frac{x_{i}^{(k)}}{\sum_{j=1}^{r}x_{j}^{(k)}},\label{eq:pi_gen_iteration_2}
\end{align}
where the iterative normalization in (\ref{eq:pi_gen_iteration_2})
only serves to avoid over- or underflows. As an initial guess we use:
\begin{equation}
\pi_{i}^{(0)}=\frac{n_{i}}{N},
\end{equation}
where $n_{i}=\sum_{j=1}^{r}n_{ij}$ and $N=\sum_{i=1}^{r}n_{i}$.
Equations (\ref{eq:pi_gen_iteration_1}-\ref{eq:pi_gen_iteration_2})
are iterated until the norm of change in $\boldsymbol{\pi}$ per iteration
is below a certain threshold (e.g. $10^{-10}$).

Applied to the specific structure of our TRAM count matrix, Equations
(\ref{eq:pi_gen_iteration_1}-\ref{eq:pi_gen_iteration_2}) become:
\begin{align}
(x_{i}^{I})^{(k)} & =\sum_{j=1}^{n}\frac{c_{ij}^{I}+c_{ji}^{I}}{\frac{n_{i}^{I}}{(\pi_{i}^{I})^{(k)}}+\frac{n_{j}^{I}}{(\pi_{j}^{I})^{(k)}}}+\sum_{J=1}^{m}\frac{b_{i}^{IJ}+b_{i}^{JI}}{\frac{n_{i}^{I}}{(\pi_{i}^{I})^{(k)}}+\frac{n_{i}^{J}}{(\pi_{i}^{J})^{(k)}}}\label{eq:pi_tram_iteration_1}\\
(\pi_{i}^{I})^{(k+1)} & =\frac{x_{i}^{(k)}}{\sum_{j=1}^{n}(x_{j}^{I})^{(k)}}.\label{eq:pi_tram_iteration_2}
\end{align}

\section{Simulation set up: Double well potential }

\label{sec:d_well} In the following we will describe the simulation
set up used for the exemplary double well potential. The potential
is given by: 
\begin{equation}
U(x)=\left\{ \begin{array}{ll}
-10+5(x+2)^{2} & \quad\text{if}\;x<-1\\
-5 x^{2} & \quad\text{if}\;x\ge-1\;\mathrm{and}\;x<0\\
-7.5 x^{2} & \quad\mathrm{if}\;x\ge0\;\mathrm{and}\;x<1\\
-15+7.5(x-2)^{2} & \quad\mathrm{if}\;x\ge1
\end{array},\right.\label{eq:double-well}
\end{equation}
with the particle position $x\in\mathbb{R}^{1}$.

In real systems the metastability of the system often limits the sampling.
To mimic this situation, the system was considered at a very low temperature,
where $\frac{1}{\beta_{0}}=1k_{B}T$. At this temperature, the probability
of finding a particle in the left well is very small: $\pi_{1}=0.008$.
The exact probabilities of each state at each temperature can be evaluated
by means of numerical integration : 
\begin{equation}
	\pi_1^{I} = \frac{\int_{x < 0}dx\exp(-u^{I}(x))}{\int_{-\infty}^{+\infty}dx\exp(-u^{I}(x))},
\end{equation}
where we use the reduced potential formulation of the main manuscript,
assuming that $u^{I}(x)=\frac{U(x)}{k_{B}T^{I}}$.
In real life problems we can only sample the system in order to obtain
estimates for their stationary probabilities or free energy differences.
A valid approach therefore is to use a particle diffusing according
to Langevin dynamics in the potential. The dynamics are given by:
\begin{equation}
m\frac{d^{2}x}{dt^{2}}=-\nabla U(x)-\gamma m\frac{dx}{dt}+\sqrt{2\gamma k_{B}Tm}R(t),\label{eqn:langevin}
\end{equation}
where $m$ is the mass of the particle and $x$ the position
coordinate of the particle. The force is given by $\nabla U(x)$,
$\gamma$ is a damping constant and $R(t)$ is a Gaussian random noise.
This is implemented in the Langevin leapfrog algorithm, where positions
and velocities are updated in alternating half time steps $dt$ \cite{VanGunsteren1988}.
For the simulations presented here, the time step was chosen to be
$dt=0.01$, $\gamma=1$ and the mass $m=1$. In addition to the single
particle evolving in the double well potential, $N$ solvent particles
were coupled to the particle evolving in $U$ of eq. (\ref{eq:double-well}),
allowing for an increase in the number of degrees of freedom of the
system. Each solvent particle $i$ will evolve in a harmonic potential
given by $U(y_{i})=y_{i}^{2}$, resulting an the total potential energy
of the system at any time to be given by: $U_{\mathrm{tot}}=U(x)+\sum_{i}U(y_{i})$.
The number of additional particles added vary between two and $50$
for different simulations carried out. In the following we will discuss
the three different simulation protocols used.

\subsection{Simulated tempering ST}

The first protocol we will briefly review is the simulated tempering
(ST) schedule employed. ST uses a single replica which diffuses in
temperature space~\cite{Marinari1992}. After $n$ simulation steps
a temperature move is attempted and accepted according to a Metropolis
criterion: 
\begin{equation}
P_{ST}(U(\mathbf{x})\beta^{I}\rightarrow U(\mathbf{x})\beta^{J})=\min[1,\exp(-U(\mathbf{x})\Delta\beta^{IJ}+\Delta g^{IJ})],\label{eq:st_accept}
\end{equation}
where $\Delta g^{IJ}$ is the difference in the weight factors at
the different temperatures. The weight factor $g^{I}$ ensures an
equal probability for sampling all temperatures with the same weight,
provided that it is defined as: 
\begin{equation}
g^{I}:=-\ln\int_{-\infty}^{+\infty}d\mathbf{x} \exp(-\beta^{I}U(\mathbf{x}))=-\ln Z^{I}=f^{I},
\end{equation}
with $Z^{I}$ being the partition function of the system at $T=T^{I}$,
as defined before. As $Z^{I}$ is one of the properties we actually
wish to estimate and is therefore not known a priori an initial guesses
will have to be made to approximate this. In the case of the simulations
presented here, an exact value was used. This is of course not possible
in simulations of actual biological interest. Therefore different
approaches for the initialization of the $g^{I}$ have been proposed
in order to make ST a viable simulation method~\cite{Rosta2010}.
A very simple way one might think of, depends on the mean potential
energies at each temperature, thus the difference in weight factors
can be expressed as: 
\begin{equation}
\Delta g^{IJ}=\frac{\langle U^{I}\rangle+\langle U^{J}\rangle}{2},
\end{equation}
as discussed in Park et al.~\cite{Park2007}.

In the case presented in the main paper, the ST simulations were carried
out in the following way: The temperature space was given by the set
of four temperatures distributed exponentially on the interval of
$\frac{1}{\beta}=[1k_{B}T,\ldots,10k_{B}T]$. Simulations were initiated
in $S_{1}$ and were allowed to change temperature with equal probability
to a higher or lower temperature after $n=100$ simulation steps.
From the choice of the the temperature spread with $2$ additional
solvent particles, the acceptance of proposed temperature moves was:
$P_{accept}=0.34\pm0.10$. At each simulation step the position of
the particle as well as the potential energy of the system were recorded.
From the single replica, stationary probabilities of being in state $S_1$
were estimated by direct counting the number of occurrences
of state 1 at the reduced temperature $k_{B}T=1$ and normalizing
these. The second estimate used was by means of the TRAM equations
and using the iterative proceedure as given by algorithm 1
of the main manuscript. Each simulation and estimation process was
repeated 100 times independently. Starting all simulations in state $S_1$
is responsible for an initial bias and leads to an overestimate of
the probability of finding the particle in state 1 in the beginning
of the simulation. This can be termed the burn-in phase. After a few
temperature cycles the simulation samples from global equilibrium,
having overcome the burn-in phase and the bias is overcome, thus ST
simulations will sample from the equilibrium distributions and will
slowly converge to it with an error proportional to $t_{\mathrm{corr}}^{-0.5}$.
All results of the ST simulation are shown in the main manuscript
in Fig. \ref{fig:fig1}.

\subsection{Parallel tempering (PT)}

The second simulation protocol used was parallel tempering (PT). Here
multiple replicas are evolved at the same time. The temperature space
is taken to be the same as the one from a ST simulation, but now we
accept two Metropolis steps simultaneously, giving rise to the following
acceptance probability for exchanging neighboring temperatures 
\begin{equation}
P_{PT}(u(\mathbf{x})^{I}\rightarrow u(\mathbf{x})^{J})=\min[1,\exp(\Delta\beta^{IJ}\Delta U(\mathbf{x})^{IJ})].\label{eq:pt_accept}
\end{equation}
This acceptance criterion ensures detailed balance and simulations
will convert to sample from the global equilibrium. Odd and even temperature
indices are alternated for the choice of neighboring pairs to be exchanged.
Exchanges are attempted at the same frequency as was done for the
ST schedule and average acceptance for the exchanges is similar
to the ST acceptance. The PT results where plotted in Fig.~\ref{fig:fig1}(E)
in the main manuscript. Estimates for the stationary probability were
additionally estimated using the MBAR equations, provided by the readily
available online implementation.

For Figs. \ref{fig:fig1}(E)+(F) of the main text, the system was only
perturbed with two solvent particles. In Fig. \ref{fig:fig1}(G)+(H),
the PT simulation had an additional 50 solvent particles and the upper
temperature bound was raised to $k_{B}T=15$ . If additional solvent
particles are added, the overlap between neighboring energy distributions
decreases. Thus perturbing the system with $50$ solvent particles
and a replica spacing of six temperatures would result in an average
acceptance of $P_{\mathrm{accept}}=0.038\pm0.08$ for exchanges. In
order to get a reasonable acceptance for this system the replica number
is increased to 20, now resulting in around every 5th exchange attempt
being accepted. This was then compared to the random swapping (RS)
protocol. Again each simulation was repeated independently 100 times.

\subsection{Random swapping (RS)}

Results of the RS protocol in a simulation with 50 solvent particles
were compared to the PT simulation as described above and shown in
Figs.~\ref{fig:fig1}(G)+(H) of the main manuscript. The
RS protocol follows the same ideas of the ST simulation, using a single
replica but instead of using a Metropolis acceptance always accepting
proposed temperature moves up or down in temperature. However, this
will drive the simulated replica out of equilibrium. Using xTRAM as
an estimation method allows the recovery of the correct stationary
probabilities from this replica none the less, due to the weakened
local equilibrium constraint. It may be necessary to adjust the lag
time $\tau$ in order to obtain the correct convergence. In the case
of the double-well potential simulations the native lag of the saving
interval of frames was sufficient to recover the correct convergence
behavior, c.f. the convergence seen in Fig. \ref{fig:fig1}(G) of the main manuscript.
In a future study the needed underlying theory for this simulation
protocol will be developed.

\section{Simulation set up: Folding potential }

Simulations for the folding potential were carried out in a very similar
fashion to those of the double well potential. The same Langevin integrator
was used. The form of the potential was altered to a $d$-dimensional
vector potential depending on a radius $r$, as given by Eq.~(\ref{eq:folding_pot})
in the main text. This time however, there were no additional solvent
particles perturbing the system. The potential was chosen in such
a way, that stationary probabilities can again be evaluated exactly.
As discussed, now the probability of being in the denatured state
is of interest which corresponds to state $S_2$ in the two-state discretization.
The temperature for which the stationary probability is evaluated
was chosen to be $\frac{1}{\beta}=1.1k_{B}T$. For the temperature
of interest, the stationary distribution of both discrete states is
given by: ${\pi}_1=0.9825$ and ${\pi}_2=0.0175$ . Again
data was generated using a ST, PT, and RS protocol in the same fashion
as discussed in section~\ref{sec:d_well} . All results obtained
from the folding potential simulations were shown in Fig.~\ref{fig:fig2}
of the main text.

\section{Simulation set up: alanine dipeptide }

Alanine dipeptide was simulated in explicit water using OpenMM~\cite{OpenMM}
with a detailed simulation setup described in the following: The force-field
was chosen to be the Amber96 all-atom force field~\cite{Kollman1996}.
$429$ water molecules using the TIP3P water model were added to the
system simulated in a cubic box with periodic boundary conditions.
For every simulation, an energy minimization and NVT-ensemble equilibration
of $100\,$ ps before production runs were carried out. For production
runs the time step was chosen to be $2\,$ fs, the saving interval
for coordinates was $0.1\,$ ps. Long range electrostatic interactions
were evaluated with Particle Mesh Ewald and a bonded cutoff of $1$
nm. All hydrogen bonds were constrained. The production run was carried
out using a Langevin integrator with a collision rate of $1\,\mathrm{ps}^{-1}$.\\

An interesting set of coordinates are the dihedral angle pair $\phi$
and $\psi$, for which a free energy surface at $T=300\, K$ can be
reconstructed. In fact the energy barrier that needs to be overcome
in order to get from an $\alpha$ helical conformation to a $\beta$
sheet arrangement of the dihedral angles is not very large. Therefore,
in order to increase the metastability of the four core states of
the dihedral angles, a von Mises potential was added to each of the
dihedral angle conformations. As presented in the main text, the von
Mises potential has the form: 
\begin{equation}
U(\delta)=\epsilon\exp\Big(\frac{\kappa\cos(\delta-\delta_{0})}{2\pi I_{0}}\Big).
\end{equation}
For each set of angular minima $\delta_{o}$ such a potential is constructed.
The positions of $\delta_{0}$ can be found in table~\ref{tab:tab_positions}.
\begin{table}
\begin{tabular}{|c|c|c|}
\hline 
Minimum  & $\phi$  & $\psi$ \tabularnewline
\hline 
I  & -150  & 150 \tabularnewline
II  & -70  & 135 \tabularnewline
III  & -150  & -65 \tabularnewline
IV  & -70  & -50 \tabularnewline
\hline 
\end{tabular}\caption{Angle minima for von Mises potential}

\label{tab:tab_positions}
\end{table}

The other factors were chosen as follows: $\epsilon=-40\,\mathrm{KJ/mol}$,
the angular deviation is $\sigma=45^{\circ}$, and $\kappa=\sigma^{-2}$,
and $I_{0}$ is the zeroth order Besselfunction. OpenMM is ideal for
a straight forward implementation of such an additional torsional
potential term~\cite{OpenMM}. Through the addition of this potential,
the energy barrier between states I and II and III and IV is enlarged
and the mixing of states is slowed, introducing additional metastability
into the system. This allows the demonstration of the superiority
of the estimator for very metastable situations. \\
 In order to setup a multi-temperature ensemble for alanine dipeptide
a series of REMD simulations were used for the production run. REMD
simulations were set up for temperatures ranging between $T=[300\, K\ldots600\, K]$.
All $33$ temperatures for the REMD simulation were spaced in such
a way as to achieve roughly equal exchange probability between adjacent
temperatures. Each replica was individually prepared at its respective
temperature, before initiating a production run. All initial configurations
belonged to state IV allowing for an initial bias which needs to be
overcome until the global equilibrium can be sampled. Accepted exchange
attempts were observed to occur around $15-20\%$ of the time. Exchanges
were attempted every 1 ps. For the REMD simulations $10$ independent
realizations each of $5\,$ ns were carried out. 

For the RS simulations not a single replica was used, but instead
13 replicas, exchanged every 10 ps with their corresponding nearest
neighbors in replica space, picking odd and even replicas in alternating
exchange moves. In the case of alanine dipeptide, there are no exactly
known stationary probabilities available for the given states, therefore
testing the validity of the RS schedule is more difficult. We found
that using a lag of $\tau=1$ ps was sufficient in generating probabilities
that were in good agreement with the REMD simulation estimates. All
results can be found in Fig~\ref{fig:free_energy} of the main
text.

\section{simulation-setup and Analysis of deca-alanine}

All-atom simulations of deca-alanine were carried out in explicit
water, using the TIP3P water model with the Amber03 forcefield in
GROMACS~\cite{Hess2008,Duan2003}. An REMD simulation with exponential
temperature spacing between 290 K and 400 K, using 24 replicas was
conducted and repeated independently 10 times, with all trajectories
starting from the same initial structure; an elongated non-helical
structure. The simulation was prepared in the following way: The initial
conformation was equilibrated with position restraints in an NVT ensemble
for each temperature individually for production runs. The REMD simulation
was carried out with the REMD option of GROMACS, with a saving interval
of 0.2 ps and replica exchanges attempted every 2 ps. A leap frog
integrator was used with a time step of 2 fs, constraining covalent bonds
with hydrogen atoms using the LINCS algorithm. For the electrostatic calculations PME was
used with a PME order of 4 and a Fourier spacing of 0.16. The NVT
ensemble was simulated using the v-rescale thermostat~\cite{Bussi2007},
coupling separately to the water and polypeptide. In all cases periodic
boundary conditions were used in a cubic box with 2609 water molecules.
For the analysis of the simulation, the temperature replicas were
detangled and sorted according to replica and not temperature. From
the all atom coordinates all combinations of $\mathrm{C}_{\alpha}$
distances were extracted, ignoring neighboring $\mathrm{C}_{\alpha}$
distances. This set of distances for all temperatures and all replicas
served as an input for a time-lagged independent component analysis
(TICA), trying to extract the most uncorrelated set of coordinates, allowing
for a dimensional reduction with the idea of retaining interesting
conformational properties with metastable barriers in between them \cite{Perez-Hernandez2013}. The TICA coordinates were calculated
with EMMA choosing three leading coordinates onto which each replica
trajectory was projected \cite{Senne2012}. On this three-dimensional TICA coordinate
trajectory, a regular spatial clustering was carried out for each
replica using a cutoff distance of 3. This resulted in 44 clusters
for each trajectory, from which all 24 discrete replicas for each
of the independent simulation runs was computed. These 44 state replica
trajectories with their corresponding potential energies and temperature
indices were used as input for the xTRAM, MBAR and direct counting
analysis, with the results displayed in Fig.~\ref{fig:deca-alanine} of the main
text. As the Amber03 forcefield is known to be overly helical, using
the state that corresponds to a helical conformation seems a good
choice for the convergence analysis, as it is highly populated at
290 K, as observed in the main manuscript. The values of the helical
population are also similar to those observed in previous force field
comparison studies looking at helicity as an order parameter
of interest \cite{Best2009}.



\end{document}